\documentclass[preprint,aps,prb,showpacs,amsmath,amssymb]{revtex4}
\usepackage{graphics}
\usepackage{epsf,graphicx,pstcol,epsfig,psfrag}
\usepackage{color}

\begin{document}

\title{ 
Static and dynamic properties of the spinless Falicov-Kimball model}

\author{K.W.~Becker$^{a}$, S.~Sykora$^{a}$, and V.~Zlatic$^{a,b}$}
\affiliation{NN
   $^{a}$Institut f\"{u}r Theoretische Physik,
   Technische Universit\"{a}t Dresden, D-01062 Dresden, Germany \\
   $^{b}$Institute of Physics, Bijenicka c.~46, P.O.B 304, 10000 
   Zagreb, Croatia
  }

\date{\today}

\begin{abstract}
The spinless Falicov-Kimball model is studied by the use of a recently developed 
projector-based renormalization method  (PRM) for many-particle Hamiltonians.
The method is used to evaluate static and dynamic quantities of the one-dimensional 
model at half-filling. To these belong the quasiparticle excitation energy 
$\tilde{\varepsilon}_k$  and the momentum distribution $n_k$ of the 
conduction electrons and spatial correlation functions of the localized electrons. 
One of the most remarkable results is the appearance of a gap in
$\tilde{\varepsilon}_k$ at the Fermi level of the order of the 
Coulomb repulsion $U$, which is accompanied by a smooth behavior for $n_k$. 
The density of states for the conduction electrons and the one-particle 
spectral functions for the localized electrons are also
discussed. In both quantities a gap opens with increasing $U$.     

\end{abstract}


\maketitle

\section{Introduction}
The Falicov-Kimball model is a widely used model to study the properties of 
interacting fermions\cite{falicov_kimball}   
and has also been used to describe the valence change transition 
in YbInCu$_4$ and similar compounds\cite{freericks.98}. 
(For a recent review see Ref.~ \onlinecite{freericks.03}).  
The model considers a lattice of localized $f$-sites, which are either 
empty or singly-occupied site, and of conduction electron sites, 
which are delocalized by the nearest-neighbor hopping. 
There are $n_f$ $f$-electrons and $n_c$ conduction electrons 
interacting by an on-site Coulomb interaction $U$. 
The two types of electrons share the common chemical potential $\mu$, 
which is always adjusted so as to keep $n_f+n_c=1$. 
The local $f$-charge is a constant of motion but thermal fluctuations 
can change the average $f$-occupation by transferring 
electrons or holes from the conduction band to $f$-states. 
The relative occupation of $f$-states and conduction states
is determined by the competition between the entropy of the $f$- and 
the  conduction states, the excitation energy of the f-states, the kinetic 
energy of band states, and the interaction energy. 

The general solution of the Falicov-Kimball model 
is not known except in infinite dimension
\cite{freericks.03,brandt.89,dongen.90}. 
In low  dimensions, the zero-temperature phase diagram 
is highly non-trivial and has attracted a lot of theoretical 
attention  (see (\onlinecite{freericks.03}) for the list of references). 
In one dimension the thermodynamic properties of the model 
are more or less understood, whereas  dynamical properties have 
been much less studied.
Kennedy and Lieb \cite{kennedy.86,lieb.86} proved that at low enough 
temperatures the half-filled Falicov-Kimball model for dimensions $d\geq 2$ 
possesses long-range order, i.e., the ions form a checkerboard pattern, 
the same as in the ground state. In the ordered phase, the lattice can be divided 
into two inter-penetrating sublattices A and B in such a way that all nearest 
neighbors of a site from sublattice A belong  to sublattice B and {\it vice versa}.  
This result holds for any value of the
Coulomb interaction $U$. For one dimension longe range order exists
only at temperature $T=0$. 

In what follows, we first describe the Hamiltonian and the projector-based 
renormalization method (PRM) which was recently introduced (Sect.~I). 
In Sect.~II the method is then applied to the one-dimensional Falicov-Kimball model 
and renormalization equations are derived for the parameters of the Hamiltonian. 
The results for static and dynamic quantities are  discussed in Sect. III and IV.  

To simplify the calculation, we consider only the case with translation symmetry. 
This excludes the discussion of physical properties which 
depend on lattice sites $i$, i.e. we do not discuss the ordered phase 
which emerges at low temperatures. However, as discussed in Sect.~V, 
we can still evaluate physical quantities which are given by averages over 
all lattice sites. Note that the PRM approach can also be extended to 
situations with long range order. 
This was shown for instance in  Ref.~(\onlinecite{sykora.05}),  
where the quantum phase transition for the spinless Holstein model 
was discussed as function of the electron-phonon coupling. 
\subsection{Model\label{FK}}
The spinless Falicov-Kimball model is described by the Hamiltonian 
\begin{eqnarray}
\label{G1}
  {\cal H} &=& 
\sum_{<i,j>} t_{ i,j}  \; 
c_i^\dagger c_j
+
\varepsilon_f\sum_i n_i +
  U \sum_i n_i n_i^c , 
\end{eqnarray}
where, $c_i^\dagger$ and $ c_i$ are the creation and 
annihilation operators for 
conduction electrons at site $i$, $t_{ i,j}$ are the hopping matrix
elements, $\varepsilon_f$ is the energy level of the localized electrons, 
and $n_i= f_i^\dagger f_i$ and $n_i^c= c_i^\dagger c_i $ are the occupation 
number operators of the localized and 
conduction electrons at site $i$. 
The summation runs over the $N$ sites of a periodic lattice, 
and $U$ is the interaction strength of the local Coulomb repulsion at sites
$i$ between conduction and localized  electrons.  
Note that $n_i$ at each site  $i$ 
commutes with the Hamiltonian. Thus, the $f$-electron occupation number
is a good quantum number, taking only two values 1 or 0, according to whether 
or not site $i$ is occupied by the localized $f$-electron.    

\subsection{Projector-based renormalization method (PRM)}
\label{PRM}

Let us first discuss a recently introduced many-particle  
method, the projector-based renormalization 
method (PRM) \cite{becker}, 
which provides an approximate solution to many-particle 
problems defined by the Hamiltonian $\mathcal{H}$,  
\begin{eqnarray}
                                      \label{G2}
  {\cal H} &=& {\cal H}_{0} + {\cal H}_{1}.
\end{eqnarray}
Here,  ${\cal H}_{0}$ is the unperturbed part with a known eigenvalue 
spectrum, 
$  {\cal H}_{0} |n\rangle = E^{(0)}_n |n\rangle$,
and $ {\cal H}_{1}$ is the perturbation which does not commute 
with ${\cal H}_{0}$ and has off-diagonal matrix elements, 
$ \langle n|{\cal H}_{1}|m\rangle\neq 0$, but no diagonal elements. 
In a usual perturbative approach one evaluates the corrections 
to the eigenstates and eigenvalues of  ${\cal H}_{0}$ due 
to ${\cal H}_{1}$.  The n-th order correction requires 
the evaluation of all matrix elements 
up to $ \langle n| {({\cal H}_{1})}^n  |m\rangle$, which are 
difficult to calculate. 

An alternative insight  is obtained by making a unitary transformation 
to a new basis which has no transition matrix elements with energy differences 
larger than some chosen cutoff $\lambda<\Lambda$. Here $\Lambda$ is the 
largest energy difference between any two eigenstates 
of ${\cal H}_0$ which are connected by ${\cal H}_1$.    
This generates a new Hamiltonian 
\begin{eqnarray}
                           \label{G4}
  {\cal H}_{\lambda} &=&
  e^{X_{\lambda}} \, {\cal H} \, e^{-X_{\lambda}} , 
\end{eqnarray}
which can be written as a sum of two terms,  
\begin{eqnarray}
                       \label{G5}
  \mathcal{H}_{\lambda} &=& \mathcal{H}_{0,\lambda} + \mathcal{H}_{1,\lambda},  
\end{eqnarray}
such that 
$\langle m| \mathcal{H}_{1,\lambda}  |n\rangle=0$ 
for $|E^{\lambda}_n-E^{\lambda}_m| > \lambda$, where 
$E^{\lambda}_n$ and $|n\rangle$ are the new eigenvalues and eigenstates 
of $\mathcal{H}_{0,\lambda}$. $\mathcal{H}_{1,\lambda}$ 
is chosen in such a way that it has no diagonal
elements with respect to  $\mathcal{H}_{0,\lambda}$.
The generator of the unitary transformation is denoted by $X_\lambda$
which has to be anti-hermitian 
$X^{\dagger}_{\lambda}=-X_{\lambda}$. Note that 
the elimination of high energy transitions may also generate new interaction
terms which have operator structures different 
from that of ${\cal H}_1$. However,  
they do not connect states with an energy separation larger than 
$\lambda$. 

In the PRM method  
the elimination of transition matrix elements is carried out 
by defining a generalized projection operator ${\bf P}_{\lambda}$ 
which removes from any operator $\mathcal{A}$ those parts which give 
rise to 'forbidden' transitions, i.e., 
\begin{eqnarray}
                                       \label{G6}
  {\bf P}_{\lambda}\mathcal{A} &=& 
  \sum_{
    \genfrac{}{}{0pt}{1}{
      \genfrac{}{}{0pt}{1}{m,n}{
        \left| E_n^{\lambda} - E_m^{\lambda} \right| \leq \lambda
      }
    }{}
  }
  |n\rangle \langle m| \, \langle n| \mathcal{A}  |m\rangle , 
\end{eqnarray}
i.e.~we retain in \eqref{G6} only the states $|m\rangle$ and $|n\rangle$ 
with $|E_n^{\lambda}  - E_m^{\lambda} |\leq \lambda$. 
The orthogonal complement of ${\bf P}_{\lambda}$ is 
$  {\bf Q}_{\lambda} = {\bf 1} - {\bf P}_{\lambda}$. Here only dyads 
$|n\rangle \langle m|$ contribute with $|E_n^{} - E_m|>\lambda$.

The generator  ${X_{\lambda}}$ 
in \eqref{G4} is determined by the 
condition 
\begin{eqnarray}   
\label{G8}
{\bf Q}_{\lambda}{\cal H}_{\lambda}=0, 
\end{eqnarray}
as can be seen by expanding the exponentials in \eqref{G4}.  
This gives 
\begin{eqnarray}
\label{G9}
 {\bf Q}_{\lambda}
\left\{
{\cal H}_{1}
+
\sum_{n=1}^\infty \frac{1}{n!}\; 
{{\bf X}_{\lambda}}^n {\cal H}
\right\}
&=& 
 0 , 
\end{eqnarray}
where  ${\bf X}_{\lambda}$ is a superoperator defined by 
${\bf X}_{\lambda} {\cal H}=[X_{\lambda}, {\cal H}]$, 
${{\bf X}_{\lambda}}^2 {\cal H}=[X_{\lambda}, [X_{\lambda}, {\cal H}]]$,  
etc. 
In lowest order,  ${\bf Q}_\lambda {\cal H}_\lambda=0 $ reduces to  
${\bf Q}_\lambda \left\{ {\cal H}_{1} -
{\bf L}_0 {X}_{\lambda}^{(1)} \right\} = 0$,
where ${\bf L}_0  {\cal A}=[{\cal H}_{0}, {\cal A}]$ 
defines the unperturbed Liouville  operator  ${\bf L}_0$,
which commutes with ${\bf Q}_{\lambda}$.  
Obviously, the condition ${\bf Q}_{\lambda}{\cal H}_{\lambda}=0$ is satisfied 
by the expression 
${X}_{\lambda}^{(1)}={\bf L}_0^{-1}{\bf Q}_{\lambda}{\cal H}_{1}$, 
i.e.,~to lowest order, ${X}_{\lambda}^{(1)}$ can be obtained from the 
decomposition of ${\cal H}_1$ into the eigenmodes of ${\bf L}_0$. 
The higher order correction terms  to ${X}_{\lambda}$ follow systematically 
from the higher order commutators of \eqref{G9}
(for details see, e.g.,  Ref.~\onlinecite{becker}). 

We see from Eq.~\eqref{G9} that all transitions in ${\cal H}_1$ with 
energy transfers between the original cutoff $\Lambda$ and the new cutoff 
$\lambda$ are eliminated in one step.  However it is more convenient 
to perform the elimination procedure step-wise,  
such that  each step reduces the cutoff energy $\lambda$ 
by a small amount $\Delta \lambda$. 
Thus, the first step removes all the transitions 
which involve energy transfers 
between (the original cutoff) $\Lambda$
and  $\Lambda- \Delta \lambda$, the subsequent steps remove all transitions 
larger than $\Lambda- 2\Delta \lambda$, $\Lambda- 3\Delta \lambda$, 
and so on. The unitary transformation for the 
step from an intermediate cutoff $\lambda$ to the 
new cutoff $\lambda - \Delta \lambda$ 
reads (in analogy to \eqref{G4}) 
\begin{eqnarray}
                            \label{G11}
  \mathcal{H}_{\lambda-\Delta\lambda} &=& 
  e^{X_{\lambda,\Delta\lambda}} 
  \, \mathcal{H}_{\lambda} \, 
  e^{-X_{\lambda,\Delta\lambda}} 
\end{eqnarray}
where $\mathcal{H}_{\lambda} = {\cal H}_{0,\lambda}+{\cal H}_{1,\lambda}$
and    
$\mathcal{H}_{\lambda-\Delta\lambda}= 
{\cal H}_{0,\lambda-\Delta \lambda}+
{\cal H}_{1,\lambda- \Delta \lambda}$.
The generator $X_{\lambda,\Delta\lambda}$ is now fixed by the condition
\begin{eqnarray}
                    \label{G12}
  {\bf Q}_{\lambda-\Delta\lambda} {\cal H}_{\lambda-\Delta\lambda} &=& 0.    
\end{eqnarray}
which specifies that ${\cal H}_{\lambda - \Delta \lambda}$ contains
no matrix elements which connect eigenstates of ${\cal H}_{0, \lambda -\Delta 
\lambda}$ with energy differences larger than $\lambda - \Delta \lambda$.

Let us assume that the operator structure of 
$\mathcal{H}_{\lambda}$ is invariant with respect to 
further unitary transformations.  Then, we can use 
Eq.~\eqref{G9} to derive difference equations for 
the $\lambda$-dependence of the coupling constants of the 
Hamiltonian. These equations connect 
the parameters of the Hamiltonian with cutoff $\lambda$ to those with cutoff 
$\lambda - \Delta \lambda$. By using a finite number of steps we can
proceed to  $\lambda\rightarrow 0$ 
and obtain a set of nonlinear equations for the renormalized parameters.  
The solution  determines the final, fully renormalized Hamiltonian 
${\cal H}_{\lambda \rightarrow 0}={\cal H}_{0,\lambda
\rightarrow 0}$, which depends on the initial parameter values of the original
model at cutoff $\Lambda$.
In this limit all transitions due to the interaction with nonzero 
 transition energies have been  
eliminated so that ${\cal H}_{1,\lambda \rightarrow 0}$ identically 
vanishes, i.e.~${\cal H}_{1,\lambda \rightarrow 0}=0$.

The underlying idea of the PRM, namely the elimination of the interaction terms 
by the unitary transformations, has been used before in the literature. 
For instance, in Ref.\onlinecite{kohn.64} 
a unitary transformation has been employed to eliminate 
off-diagonal matrix elements of the Coulomb interaction 
in order to study ground state properties of the Hubbard model. 
Similarly, in the 
Schrieffer-Wolff transformation \cite{schrieffer-wolff}, which  maps the 
Anderson to the Kondo model, 
matrix elements connecting different charge configurations 
of magnetic ions are eliminated also by use of a unitary transformation.
Note however that in the  
PRM approach unitary transformations are performed in small steps   
in contrast to earlier applications. Note also that the approach removes  
high energy transitions but does not decimate the Hilbert space. 
This  is different to the poor man's 
scaling\cite{anderson} which removes high energy states. 
It should also be mentioned that the 
PRM resembles the previous flow equation 
method \cite{wegner} by Wegner and the similarity 
renormalization approach\cite{wilson}
by Glazek and Wilson which can be considered as continuous versions of the 
PRM method. 

\section{Renormalization of the Falicov-Kimball model}

In order to apply the PRM we express the Falicov-Kimball 
interaction in terms of the fluctuations with respect to the 
thermal averages of $n_i^c$ and $n_i$, 
and write the diagonal and off-diagonal part of ${\cal H}$ as 
\begin{eqnarray}
                          \label{G13}
{\cal H}_0 &=& \sum_i \big( \varepsilon_f + U \big< n_i^c \big> \big)\;
n_i +
\sum_k \big( \varepsilon_k + U \big< n_i \big> \big) \;
c_{k}^\dagger c_k - UN \big<n_i\big>\; \big< n_i^c \big> ,
\end{eqnarray} 
and 
 \begin{eqnarray}
                       \label{G14}
 {\cal H}_1 &=& \frac{U}{N} \sum_{i k {k'}}\;
e^{-i(k - {k'}){R}_i}\;
\delta n_i\; \delta (c_k^\dagger c_{k'}),  
\end{eqnarray}
where $\delta {n_i^c}= {n_i^c}- \big< {n_i^c}\big>$, 
$\delta {n_i}= {n_i}- \big< {n_i}\big>$,  
$\big< {\cdots}\big>$ denotes the thermal averaging with the full 
Hamiltonian ${\cal H}$, and 
$ 
c_{k}^\dagger =
({1}/{\sqrt N}) \sum_i e^{ik{R}_i}c_{i}^\dagger 
$
is the Fourier transform of $c_i^\dagger$.  
Of course, due to the assumed translation invariance 
$\langle n_i \rangle$ is equivalent to 
$(1/N)\sum_i  \langle n_i \rangle$, where $N$ is the number of lattice sites.
The sum  in \eqref{G14} is restricted to $k\neq k'$, 
since the diagonal part of the Coulomb repulsion is included in ${\cal H}_0$. 
From translation invariance it follows 
$\delta(c_k^\dagger c_{k'}) =  c_k^\dagger c_{k'} 
- \big< c_k^\dagger c_k\big> \delta_{k, k'}$.
In principle, the average values in \eqref{G13} can be defined 
with an arbitrary ensemble since Eq.~\eqref{G1} and  
Eqs.~\eqref{G13}, \eqref{G14} are equivalent. 
It has turned out that the best 
choice is to evaluate the averages with respect to the 
original full   
Hamiltonian ${\cal H}= {\cal H}_{\Lambda}$, which can be done by 
the procedure explained below. 

The perturbation ${\cal H}_1$ causes transitions  between the 
eigenstates of ${\cal H}_0$ by creating electron-hole pairs 
in the conduction band, whereas the number of $f$-electrons is conserved 
at each site. Using 
${\bf L}_0 \; \delta n_i \; \delta(c_k^\dagger c_{k'}) 
= (\varepsilon_k - \varepsilon_{k'}) \;
\delta n_i \; \delta(c_k^\dagger c_{k'})$,  
${\cal H}_1$ can be decomposed into a sum of 
eigenmodes of the unperturbed Liouville operator. 
The lowest order solution of the generator of the unitary
transformation is 
$X_\lambda^{(1)}={\bf L}_0^{-1} {\bf Q}_{\lambda} {\cal H}_1 $, 
which is given by 
\begin{eqnarray}
                   \label{G16}
X_\lambda^{(1)} 
&=& 
\frac{U}{N} \sum_{i k k'}\;
\frac{e^{i(k - k'){R}_i}}
{\varepsilon_k - \varepsilon_{k'}}\;
\Theta({|\varepsilon_k - \varepsilon_{k'}|}- \lambda)\;
\delta n_i\; c_k^\dagger c_{k'}. 
\end{eqnarray}
Here 
$\Theta({|\varepsilon_k - \varepsilon_{k'}|}- \lambda)$
is the projection operator which removes from ${\cal H}_1$
all transitions with energy 
transfers larger than $\lambda$. Note that the
same operator form also appears in the non-perturbative calculation
below. 

\subsection{Renormalized Hamiltonian\label{renormalizedH}}  
The unitary transformation leads to the renormalized Hamiltonian 
${\cal H}_\lambda = e^{X_\lambda} {\cal H}  e^{-X_\lambda}$,  
which can be written as 
${\cal H}_\lambda ={\cal  H}_{0,\lambda} + {\cal H}_{1,\lambda}$, 
where ${\cal H}_{1,\lambda}$ gives rise to the transitions between the 
eigenstates of ${\cal  H}_{0,\lambda}$ but has no diagonal elements.  
The choice of ${\cal  H}_{0,\lambda}$ is not unique,  
and each particular case requires a physical intuition.
But in any case it has to be such that the thermal 
averages which appear in the procedure can be evaluated. 
For the Falicov-Kimball model, the choice which emphasizes the 
weak coupling limit, is 
\begin{eqnarray}
                   \label{G17}
{\cal H}_{0,\lambda} &=&  \varepsilon_{f,\lambda} \sum_i \delta n_i
+
\sum_{i \neq j} g_{ij, \lambda} \; \delta n_i \; \delta n_j
+ \sum_k \varepsilon_{k,\lambda} \; c_k^\dagger c_k 
+ E_\lambda ,\\
&& \nonumber \\ 
{\cal H}_{1,\lambda} &=& 
 \frac{U}{N} \sum_{i k k'} 
\Theta (\lambda -|\varepsilon_{k, \lambda} -\varepsilon_{k', \lambda}|) \; 
e^{i(k - k'){R}_i}\;
\big( \delta n_i\; \delta (c_k^\dagger c_{k'})
\big) , 
                 \label{G18}
\end{eqnarray}
where 
$ \varepsilon_{f,\lambda}$, $ g_{ij, \lambda}$, and
$\varepsilon_{k,\lambda}$
are the renormalized parameters which are calculated below.
Note that the interaction parameter $U$ will not 
change in the renormalization procedure below
so that a $\lambda$-dependence of $U$  is not considered 
from the beginning.
The second term in \eqref{G17} is a new density-density interaction between
the localized electrons which is generated during the renormalization 
procedure.  
Additional higher order operator terms, which are also generated 
by the renormalization, are neglected. 
The original model ($\lambda = \Lambda$) is defined by the 
initial condition, according to \eqref{G13}, \eqref{G14}  
\begin{eqnarray}
\label{G19}
&& \varepsilon_{f, \Lambda} = \varepsilon_f + U \big< n_i^c \big>
\hspace*{2cm}
\varepsilon_{k, \Lambda} = 
\varepsilon_k + U \big< n_i\big>  \\
&& g_{ij, \Lambda} =0 \hspace*{3.7cm}
E_{\Lambda} = -N\big<n_i \big>\; ( \varepsilon_f +
U \; \big< n_i^c \big>) . \nonumber 
\end{eqnarray}  

Once we have made the ansatz \eqref{G17} and \eqref{G18} for the 
operator structure of  ${\cal H}_{0,\lambda} $ and 
${\cal H}_{1,\lambda}$,  we 
reduce the  cutoff to $\lambda -\Delta  \lambda$
to  find the new effective Hamiltonian 
\begin{eqnarray}
\label{G20}
{\cal H}_{\lambda -\Delta  \lambda} &=&
e^{X_{\lambda, \Delta \lambda}} {\cal H}_\lambda 
e^{-X_{\lambda, \Delta \lambda}} 
={\cal H}_{0,\lambda -\Delta  \lambda}
+{\cal H}_{1,\lambda -\Delta  \lambda } 
\end{eqnarray}
with renormalized parameters. The new Hamiltonian is such 
that ${\cal H}_{1,\lambda -\Delta  \lambda }$ 
does not give rise to transitions with energy transfers  
larger than $(\lambda -\Delta \lambda)$ (with respect to 
the new unperturbed part  ${\cal H}_{0,\lambda -\Delta  \lambda }$). 
${\cal H}_{\lambda - \Delta \lambda}$ should have the same operator
structure as ${\cal H}_{\lambda}$, however with renormalized 
parameters 
\begin{eqnarray}
                            \label{G31}
 {\cal H}_{\lambda- \Delta \lambda} &=&  
\varepsilon_{f,(\lambda- \Delta \lambda)} \sum_i \delta n_i
+
\sum_{i \neq j} g_{ij, (\lambda - \Delta \lambda)} \;
\delta n_i \; \delta n_j 
+
\sum_k \varepsilon_{k,(\lambda - \Delta \lambda)} 
\; c_k^\dagger c_k + E_{(\lambda - \Delta \lambda)}
 \\ 
&+&
 \frac{U}{N} \sum_{i k k'} \Theta(\lambda - \Delta \lambda
 -|\varepsilon_{k, \lambda  
- \Delta \lambda} - \varepsilon_{k', \lambda - \Delta \lambda}|) \; 
e^{i(k - k'){R}_i}\;
\big( \delta n_i\; \delta (c_k^\dagger c_{k'})
\big) . 
\nonumber 
 \end{eqnarray}

\subsection{Unitary transformation\label{generator}}
For the explicit evaluation of the unitary transformation \eqref{G20}
we need the generator $X_{\lambda, \Delta \lambda}$. We make the 
following ansatz 
\begin{eqnarray}
                                 \label{G22}%
&& X_{\lambda, \Delta \lambda} = \frac{1}{N} \sum_{k, k',i}
A_{k, k'}^{\lambda,\Delta\lambda} \;
\Theta_{kk' }^{\lambda,\Delta\lambda}
\;
 e^{-i(k- k'){R}_i }\;
\delta n_i \; 
\delta( c_k^\dagger c_{k'}) . 
\end{eqnarray} 
Here $\Theta_{kk' }^{\lambda,\Delta\lambda}$ is a product 
of $\Theta$-functions 
\begin{eqnarray}
                                \label{G23}
\Theta_{kk' }^{\lambda,\Delta\lambda} &=&
\Theta(\lambda - |\varepsilon_{k, \lambda} -
 \varepsilon_{k', \lambda}|) 
\times 
\Theta \big[|\varepsilon_{k, (\lambda- \Delta \lambda)} -
 \varepsilon_{k', (\lambda - \Delta \lambda)}| - 
(\lambda - \Delta \lambda) \big] , 
\end{eqnarray}
which project all the operators in $X_{\lambda, \Delta \lambda}$ 
'on the shell' between $\lambda$ and $\lambda- \Delta \lambda$.
More precisely, the energy differences 
$|\varepsilon_{k, \lambda} -\varepsilon_{k', \lambda}|$
and $|\varepsilon_{k, (\lambda- \Delta \lambda)} -
 \varepsilon_{k', (\lambda - \Delta \lambda)}|$ in \eqref{G23} 
refer  to the two different Hamiltonians 
${\cal H}_{\lambda}$ and ${\cal H}_{\lambda- \Delta \lambda}$.
Therefore, the two $\Theta$-functions in \eqref{G23} take into account that 
${\cal H}_\lambda$ possesses only 
transition elements with  energy transfer $|\varepsilon_{k, \lambda} -
\varepsilon_{k', \lambda}| < \lambda$, whereas in
${\cal H}_{\lambda- \Delta \lambda}$ no transitions with 
$|\varepsilon_{k, (\lambda- \Delta \lambda)} -
 \varepsilon_{k', (\lambda - \Delta \lambda)}| >
\lambda - \Delta \lambda$ are allowed.  
The coefficients  $A_{k, k'}^{\lambda,\Delta\lambda}$ 
are determined below, using the condition 
${\bf Q}_{\lambda-\Delta\lambda} {\cal H}_{\lambda-\Delta\lambda} = 0$. 
Again, the form of  $X_{\lambda, \Delta \lambda}$ 
is suggested by its lowest order expression according to    
\eqref{G16} (compare (\onlinecite{becker})). 
The unitarity of the transformation requires 
$X_{\lambda,\Delta\lambda}=-X_{\lambda,\Delta\lambda}^\dagger$, 
so that $A_{k, k'}^{\lambda,\Delta\lambda}$ has to be 
antisymmetric with respect to the interchange of $k$ and $k'$.   
Expanding the exponentials in \eqref{G20} 
the transformation can be written as,  
\begin{eqnarray}
                           \label{G25}
{\cal H}_{\lambda - \Delta \lambda} &=&
e^{X_{\lambda, \Delta \lambda}} {\cal H}_{\lambda}
e^{-X_{\lambda, \Delta \lambda}} 
= {\cal H}_{\lambda}
+
\sum_{n=1}^\infty \frac{1}{n!}\; 
{\bf X}_{\lambda, \Delta \lambda}^n {\cal H}_{\lambda}
\end{eqnarray}
where 
${\bf X}_{\lambda, \Delta \lambda} {\cal H}_{\lambda} = 
[X_{\lambda, \Delta \lambda}, {\cal H}_{\lambda}]$, 
${\bf X}_{\lambda, \Delta \lambda}^2 {\cal H}_{\lambda} = 
[X_{\lambda, \Delta \lambda}, [X_{\lambda, \Delta \lambda}, 
{\cal H}_{\lambda}]]$, etc. The commutators are evaluated 
in Appendix A. 
The basic approximation of the PRM approach is to replace 
some of the operators generated by $X_{\lambda,\Delta\lambda}$ 
by ensemble averages, so as to keep the structure of 
${\cal H}_\lambda$ invariant during the renormalization procedure. 
This additional factorization enables us  
to resum the series in \eqref{G25}.

\subsection{Renormalization of the coupling constants\label{couplings}}
By comparing the respective operators in \eqref{G31} with those 
obtained from \eqref{G25} one finds renormalization equations 
for the parameters of ${\cal H}_\lambda$. First, from  
the prefactors of $\delta n_i$ one obtains
\begin{eqnarray}
                        \label{G33}
\varepsilon_{f, (\lambda - \Delta \lambda)}
&=&  \varepsilon_{f, \lambda} +  
\frac{(1- 2\big< n_i\big>)}{2N} \sum_{k k'} 
\Theta_{k, k'}^{\lambda, \Delta \lambda}     \;
(\varepsilon_{k', \lambda}-
\varepsilon_{k, \lambda})\;
\frac{(1- \cos v_{k k'}^{\lambda})}{C_\rho(k -k')} \;
\big< c_k^\dagger c_k \big> \\
&+& \frac{U(1-2\big< n_i \big>)}{N \sqrt{N}}
\sum_{k k'} 
\Theta_{k, k'}^{\lambda, \Delta \lambda}
\frac{\sin v_{k k'}^{\lambda}}{\sqrt{C_\rho(k-k')}}
\; \big< c_k^\dagger c_k \big> .
\nonumber
\end{eqnarray}
Similarly, by comparing 
the coefficients of $c_k^\dagger c_k$ one finds  
\begin{eqnarray}
                          \label{G36}
\varepsilon_{k,(\lambda - \Delta \lambda)} &=&
\varepsilon_{k,\lambda} 
+ \frac{1}{2} \sum_{k'} 
\Theta_{k, k'}^{\lambda, \Delta \lambda}     \;
(\varepsilon_{k',\lambda}- \varepsilon_{k, \lambda} )\;
( 1- \cos v_{k k'}^{\lambda}) \\
&+& \frac{U}{\sqrt{N}} \sum_{k'}
\Theta_{k, k'}^{\lambda, \Delta \lambda}     \;
 \sqrt{C_\rho(k- k')}\; \sin v_{k k'}^{\lambda} .
\nonumber
\end{eqnarray}
The renormalization equation for the interaction parameter 
$g_{ij,\lambda}$ between $f$-electrons is best expressed  
in terms of  its Fourier transform $g_{{q}, \lambda}$, 
\begin{eqnarray}
                    \label{G39}
g_{{q}, (\lambda - \Delta \lambda)} &=& 
g_{{q}, \lambda } 
+ \frac{1}{4} 
\sum_k  \frac{\big<c_k^\dagger c_k\big>
-\big<c_{k+{q}}^\dagger c_{k+{q}} \big>}
{C_\rho({q})}\; 
\Theta_{k, k+{q}}^{\lambda, \Delta \lambda}\;
(\varepsilon_{k+{q}, \lambda} - 
\varepsilon_{k, \lambda})\; 
(1- \cos v_{k, k+{q} }^\lambda )
\nonumber \\ 
 &+& \frac{U}{2\sqrt{N}} \sum_k
\frac{\big<c_k^\dagger c_k\big> -
\big<c_{k+{q}}^\dagger c_{k+{q}}\big>}
{\sqrt{C_\rho({q})}}\;
 \Theta_{k, k +{q}}^{\lambda,
  \Delta \lambda}\;
\sin v_{k, k+{q}}^\lambda \\
&-&  \frac{1}{N} \sum_{q'} (\cdots) \nonumber
\end{eqnarray}
where $(\cdots)$ denotes the second and third term on the r.h.s.~with
the wave vector ${q}$ replaced by ${q'}$.  In this way, 
the exact sum rule $\sum_{q} g_{{q},\lambda}
=0$ is fulfilled which guarantees that only  sites 
with $i \neq j$ contribute to $g_{ij}$. 
Finally, also the renormalization equation for 
the energy shift $E_\lambda$ can be obtained which will not be given
explicitly. The other new quantities in Eqs.~\eqref{G33} to \eqref{G39} 
are defined as follows
\begin{eqnarray} 
                        \label{G28}
&& v_{k k'}^{\lambda} = 2 \sqrt{\frac{C_\rho(k- k')}{N}}
\; A_{k, k'}^{\lambda,\Delta\lambda},
\hspace*{1cm} 
C_\rho({q}) = \frac{1}{N} \sum_{ij} e^{i{q}({R}_i- 
{R}_j)} \big< \delta n_i \delta n_j \big>. \\
&& \nonumber
\end{eqnarray} 
Finally we have to determine the coefficients 
$A_{k, k'}^{\lambda,\Delta\lambda}$ of the unitary 
transformation. They follow from the condition 
${\bf Q}_{\lambda-\Delta\lambda} {\cal H}_{\lambda-\Delta\lambda}=0$,  
which removes from ${\cal H}_{\lambda-\Delta\lambda}$ 
all the operators giving rise to the transitions with energy 
transfers larger than $\lambda-\Delta\lambda$. One finds
\begin{eqnarray}
                                    \label{G30}
\Theta_{kk' }^{\lambda,\Delta\lambda} 
\; A_{k, k'}^{\lambda,\Delta\lambda}&=&
\Theta_{kk' }^{\lambda,\Delta\lambda} \; \frac{1}{2} \;
\sqrt{\frac{N}{C_\rho(k - k')}}\;
\arctan\Big( 2 \sqrt{\frac{C_\rho(k- k')}{N}}\;
\frac{U}{\varepsilon_{k, \lambda}
-\varepsilon_{k', \lambda}}
\Big) .
\end{eqnarray}

Note that the $A_{k, k'}^{\lambda,\Delta\lambda}$ 
depend on the parameters at cutoff $\lambda$ as well as from the reduced 
cutoff ${\lambda-\Delta\lambda}$. 
The presence of $\Theta_{kk' }^{\lambda,\Delta\lambda} $  
on both sides of Eq.\eqref{G30} 
means that  $A_{k, k'}^{\lambda,\Delta\lambda}$ has to be 
specified only for those values of $k$ and $ k'$ 
which are 'on the shell' defined by the condition  
$\Theta_{kk'}^{\lambda, \Delta \lambda}=1$.   
'Off-the-shell'  coefficients 
can take any value and, in what follows, we use 
$A_{k, k'}^{\lambda,\Delta\lambda}=0$ for 
$\Theta_{kk'}^{\lambda, \Delta \lambda}=0$.   
Note that 
expression  \eqref{G30} includes  $U$ to all orders. However,  
in the thermodynamic limit 
$N \rightarrow \infty$, the coefficients 
$A_{k, k'}^{\lambda,\Delta \lambda}$ become 
linear in $U$ since all higher order terms vanish for $N\rightarrow \infty$. 
In the actual numerical evaluation of the renormalization equations 
on a lattice of finite 
size $N$ the excitation energies 
($\varepsilon_{k',\lambda}-\varepsilon_{k,\lambda}$) 
may become very small for $\lambda \rightarrow 0$ so that  the 
expansion of $A_{k, k'}^{\lambda,\Delta \lambda}$ to 
linear order in $U$ breaks down. 
In this case, the full expression \eqref{G30}  has to be taken. 

By help of \eqref{G30} one finds for the parameters $v_{k k'}^\lambda$ 
of \eqref{G28} 
\begin{eqnarray} 
\label{G35}
 \sin v_{k k'}^\lambda &=& 2 U \sqrt{\frac{C_\rho(k- k')}{N}}
\frac{\mbox{sign} \ (\varepsilon_{k,\lambda} 
- \varepsilon_{k',\lambda}) }
{ \sqrt{(\varepsilon_{k, \lambda} -
\varepsilon_{k', \lambda})^2 
+4U^2 C_\rho(k- k')/{N} }} 
\nonumber \\
 \cos v_{k k'}^\lambda &=& 
\frac{|\varepsilon_{k, \lambda} -\varepsilon_{k', \lambda}|}
{ \sqrt{(\varepsilon_{k, \lambda} -\varepsilon_{k', \lambda})^2 
+4U^2C_\rho(k- k')/{N}}} \ .
\end{eqnarray}

\vspace*{1cm}

Equations \eqref{G33} to \eqref{G39} represent the final
renormalization equations for the parameters  of the Hamiltonian 
as the energy cutoff is reduced from 
$\lambda$ to  $\lambda - \Delta \lambda$. 
The overall renormalization starts from the original 
cutoff $\lambda= \Lambda$, 
where the initial parameters are given by \eqref{G19}. For the 
parameters $ \sin v_{k k'}^\lambda$,   $ \cos v_{k k'}^\lambda$, 
and $C_\rho(k- k')$, which contain expectation values, 
an appropriate choice is taken at cutoff $\Lambda$. 
Using these quantities we obtain from the renormalization 
equations \eqref{G33} to \eqref{G39}  
the renormalized parameters at the smaller cutoff $\Lambda-\Delta\lambda$. 
The procedure is repeated until the cutoff is reduced to zero. 
At $\lambda =0$, we obtain the completely renormalized Hamiltonian 
$\tilde{\cal H}:= {\cal H}_{(\lambda=0)}$, which describes an effectively 
decoupled system of free conduction electrons and interacting 
localized electrons, 
\begin{eqnarray}
                      \label{G40}
\tilde{\cal H} &=& 
\tilde{\cal H}_c + \tilde{\cal H}_{f} + \tilde{E} , 
\end{eqnarray}
where 
\begin{eqnarray}
                     \label{G41}
\tilde{\cal H}_c &=& 
\sum_k \tilde{\varepsilon}_k\;
c_k^\dagger c_k, 
\hspace*{1cm}  \mbox{and} \hspace*{1cm} 
\tilde{\cal H}_{f} 
=\tilde{\varepsilon}_f \sum_i \delta n_i + \sum_{i \neq j} \tilde{g}_{ij}\;
\delta n_i \delta n_j .
\end{eqnarray}
The fully renormalized parameters are 
defined as $\tilde{\varepsilon}_k= 
\varepsilon_{k, (\lambda=0)},  \tilde{\varepsilon}_f= 
\varepsilon_{f,(\lambda=0)}, \tilde{g}_{ij} = g_{ij,(\lambda=0)}$
and $\tilde{E}= E_{(\lambda=0)}$. 
Note that for $\lambda \rightarrow 0$
the interaction ${\cal H}_{1,\lambda}$ has completely 
vanished due to the $\Theta$-function in 
\eqref{G18}. 
Using $\tilde{\cal H}$ we next can recalculate the statistical averages 
$\big< c_k^\dagger c_k\big>$, $C_\rho(k- k')$, 
etc., which appear in Eqs. \eqref{G33} -- \eqref{G39}, to find new 
values for the renormalized coupling parameters. This procedure is repeated
until the self-consistency is reached. 
Note that the Hamiltonian $\tilde{\cal H}_f$ for the localized 
electrons is usually denoted as the lattice gas model.
Due to the dependence of $\tilde{g}_{ij}$ on 
$i-j$, the interaction in $\tilde{\cal H}_f$ 
is not necessarily restricted to nearest neighbors.

Let us add two side remarks: 
The first one concerns the validity of the present approach.  As is obvious
from the renormalization procedure in Sect.~II we are here dealing with a 
weak-coupling approach. The Coulomb interaction $U$ is taken as the 
'small' quantity which is successively integrated out in the procedure. 
However in each renormalization step not only the lowest order in $U$ is
included. Instead, certain renormalization 
contributions up to infinite order in $U$ are taken into account. 
Therefore, we expect the present approach to be valid for   
$U$ values which are less or at most of the order of the hopping term $t$. 
For larger values of $U$ either newly generated renormalization terms would
have to be added. Or, more appropriate for large $U$ 
would be to take the dominant $U$ term as part of a new 'unperturbed' 
Hamiltonian ${\cal H}_0$ and the hopping term as part of the new perturbation 
${\cal H}_1$.      

The second remark applies to the optimal choice of the 
renormalization interval  
$\Delta\lambda$.  It is best determined by the size of the system, i.e., 
by the requirement that the 'energy shell' contains only a few 
$k$-points. Thus, the number of $\Delta \lambda$ intervals should be 
of the order of the lattice size $N$. For a larger number of intervals the
result would be the same since more steps occur 
without renormalization since no transition
energies fit into the interval. On the other hand a smaller 
number of $\Delta \lambda$ intervals leads to a quite large   
renormalization in some steps since many $k$-points could contribute 
in these intervals. 
In practise, for a given $N$ we have increased the number of intervals until 
the renormalization does not change any more.

\section{Evaluation of Static Expectation values}
The renormalization equations depend on expectation values 
$\big< n_i\big>$, $\big< n_i n_j \big>$ and $\big<c_k^\dagger 
c_k \big>$ which have been defined with 
the full Hamiltonian ${\cal H}$. Since thermal averages are invariant with 
respect to a unitary transformations, we can do the averaging 
with the renormalized Hamiltonian $\tilde{\cal H}$, provided we 
also transform the operators,  
\begin{eqnarray}
                      \label{G42}
\big< {\cal A} \big> &=& \big< {\cal A}(\lambda)  \big>_{{\cal H}_\lambda}
=  \big< {\cal A}(\lambda \rightarrow 0)  \big>_{\tilde{{\cal H}}} ,
\hspace*{1cm}
{\cal A}(\lambda)= e^{X_{\lambda}} {\cal A}  e^{-X_{\lambda}}. \\
&& \nonumber
\end{eqnarray}

To evaluate $\big< c_k^\dagger c_k \big>$ one 
needs  the renormalization equations for the operators 
$(c_k^\dagger c_k)(\lambda)= c_k^\dagger(\lambda) \;
c_k(\lambda)$, where 
 $c_k(\lambda)
 =
 e^{X_{\lambda, \Delta \lambda}}c_k
e^{-X_{\lambda, \Delta \lambda}}$.
For  $c_k(\lambda)$ we use the following ansatz, 
\begin{eqnarray}
                                \label{G45}
c_k(\lambda) &=& \alpha_{k,\lambda} c_k
+   \frac{1}{N} \sum_{k',i} \beta_{k,k',\lambda}\; 
e^{i(k -k'){R}_i} \; \delta n_i c_{k'} , 
\end{eqnarray}
where $\alpha_{k,\lambda}$ and $\beta_{k,k',\lambda}$ 
are $\lambda$-dependent coefficients. The operator structure 
of $c_k(\lambda)$ in \eqref{G45} is again taken from the 
first order expansion in $U$. 
All higher order  terms are neglected. 
The renormalization equations for the new parameters 
can be derived in close analogy to the renormalization 
of ${\cal H}_\lambda$. One finds, 
 \begin{eqnarray}
                                   \label{G46}
\alpha_{k, (\lambda -\Delta \lambda)} &=&
 \alpha_{k, \lambda} 
 - \alpha_{k, \lambda} 
\sum_{k'} \big[ 1 - \cos\big(A_{k k'}^{\lambda,\Delta \lambda}
\sqrt\frac{C_\rho(k-k')}{N} \big)\big] \;
\Theta_{kk'}^{\lambda, \Delta \lambda} \\
&+&
\frac{1}{\sqrt{N}} \sum_{k'}
\beta_{k k', \lambda} \;
\Theta_{kk'}^{\lambda, \Delta \lambda}
\sqrt{C_\rho(k -k')}
\sin\big(A_{k k'}^{\lambda,\Delta \lambda}
\sqrt\frac{C_\rho(k-k')}{N} \big)
\nonumber 
\end{eqnarray}
and 
\begin{eqnarray}
                                  \label{G47}
\beta_{kk', (\lambda -\Delta \lambda)} &=&
 \beta_{kk', \lambda} 
 - \alpha_{k, \lambda} \;
\sqrt{\frac{N}{C_\rho({k-k'})}}
\sin\big(A_{k k'}^{\lambda,\Delta \lambda}
\sqrt\frac{C_\rho(k-k')}{N} \big)\big] \;
\Theta_{kk'}^{\lambda, \Delta \lambda}
\nonumber \\
&-&
 \beta_{kk', \lambda} \; \big( 1 -
\cos\big[A_{k k'}^{\lambda,\Delta \lambda}
\sqrt\frac{C_\rho(k-k')}{N} \big] \big) \;
\Theta_{kk'}^{\lambda, \Delta \lambda}  , 
\end{eqnarray}
where 
$A_{k, k'}^{\lambda,\Delta\lambda}$
and
$C_\rho(k-k')$ are defined by Eq.~\eqref{G30} 
and \eqref{G28}, respectively. 
The initial values are again those of the original 
model ($\lambda = \Lambda$), $ \alpha_{k, \Lambda} = 1, 
\beta_{k k', \Lambda} =0$.
Eqs.~\eqref{G45} -\eqref{G47}
determine the transformation behavior of $c_k(\lambda)$. 
By the use of \eqref{G45} we finally obtain 
\begin{eqnarray}
                        \label{G49}
n_k= \big< c_k^\dagger c_k\big> &=& 
|\tilde{\alpha}_{k}|^2 
\big< c_k^\dagger c_k\big>_{\tilde{\cal H}_c}
+
\frac{1}{N} \sum_{k'} |\tilde{\beta}_{k,k'}|^2 
C_\rho(k-k')
\big< c_{k'}^\dagger c_{k'}\big>_{\tilde{\cal H}_c} , 
\end{eqnarray}
where 
$ \tilde{\alpha}_{k} =  \alpha_{k, (\lambda=0)}$
and 
$\tilde{\beta}_{k,k'} = 
\beta_{k k', (\lambda=0)}$,  
and $\big< c_k^\dagger c_k\big>_{\tilde{\cal H}_c}$
is the Fermi function formed with
$\tilde{\cal H}_c$, 
$ \big< c_k^\dagger c_k\big>_{\tilde{\cal H}_c}
= ({e^{\beta \tilde{\varepsilon}_k}+1})^{-1}$. \\

The occupation number operators $n_i$ 
of the localized electrons commute with $X_\lambda$. 
Thus $n_i$ will not be renormalized, $n_i(\lambda) = n_i$, 
$(n_i n_j) (\lambda) = n_i n_j$, i.e.,
\begin{eqnarray}
                                \label{G52}
n_f=  \big< n_i \big> &=& \big< n_i \big>_{\tilde{{\cal H}_f}} \hspace*{1cm}
\big< \delta n_i \delta n_j\big> = 
\big< \delta n_i \delta n_j \big>_{\tilde{{\cal H}_f}}.
\end{eqnarray}
Here the expectation values are formed with  
the lattice gas model $\tilde{\cal H}_f$ since the conduction 
and the localized part of the renormalized 
Hamiltonian $\tilde{\cal H}$ are decoupled. 
The expectation values in \eqref{G52} can be found by direct numerical 
evaluation in the local occupation number representation.

\section{Dynamic correlation functions\label{dynamical properties}}
Dynamical quantities can be evaluated by the PRM approach along the same
lines. In what follows we shall 
discuss the spectral functions of the conduction and 
localized electrons. 
The one-particle spectral functions of  conduction electrons 
$A_k^+(\omega)$ and  $A_k^-(\omega)$ 
are defined as 
\begin{eqnarray}
                              \label{G56}
A_k^+(\omega) &=& \frac{1}{2\pi} \int_{-\infty}^{\infty}
\big< c_{k } (t) \;c_{k}^\dagger
\big> \; e^{i\omega t} dt\,, \hspace*{1cm}
A_k^-(\omega) = \frac{1}{2\pi} \int_{-\infty}^{\infty}
\big< c_{k }^\dagger  \;c_{k}(t) 
\big> \; e^{i\omega t} dt\, ,     
\end{eqnarray}
where  $A_k^+(\omega)$ describes the creation of an electron 
$k$ at time zero and its annihilation at time $t$ whereas in
$A_k^-(\omega)$ first an electron is annihilated. 
It is well-known that $A_k^+(\omega)$ 
and  $A_k^-(\omega)$ can be measured 
by inverse photoemission (IPE) and by photoemission (PE) experiments. 
To evaluate $A_k^+(\omega)$ and $A_k^-(\omega)$ we again 
exploit the property \eqref{G42} of expectation values and the 
ansatz \eqref{G45} for $c_k(\lambda)$ and 
$c_k^\dagger(\lambda)$. We find 
\begin{eqnarray}
                             \label{G58}
A_k^+(\omega) &=& |\tilde{\alpha}_k|^2 \; 
(1 - \big< c_k^\dagger c_k\big>_{\tilde{\cal H}_c})\;
\delta(\omega -\tilde{\varepsilon}_k) 
+ \frac{1}{N} \sum_{k'} |\tilde{\beta}_{k k'}|^2\;
C_\rho(k-k')(1- 
\big< c_{k'}^\dagger c_{k'} \big>_{\tilde{\cal H}_c})\;
\delta (\omega - \tilde{\varepsilon}_{k'})  \nonumber \\
&& \\
A_k^-(\omega) &=& |\tilde{\alpha}_k|^2 \; 
\big< c_k^\dagger c_k\big>_{\tilde{\cal H}_c}\;
\delta(\omega -\tilde{\varepsilon}_k) 
+ \frac{1}{N} \sum_{k'} |\tilde{\beta}_{k k'}|^2\;
C_\rho(k-k')\;
\big< c_{k'}^\dagger c_{k'} \big>_{\tilde{\cal H}_c}\;
\delta (\omega - \tilde{\varepsilon}_{k'})  \nonumber 
\end{eqnarray}
where $\big< c_{k'}^\dagger c_{k'} \big>_{\tilde{\cal H}_c}$ 
is the Fermi distribution. Note that in \eqref{G58}
we have already reduced the cutoff to $\lambda \rightarrow 0$.
The 
first term in  $A_k^+(\omega)$ and $A_k^-(\omega)$ describes 
the coherent one-electron excitation 
which is the only excitation in a free electron gas model. 
The remaining contributions are incoherent excitations
due to the coupling of $c_k$ 
to different conduction electrons with wave 
vectors $k'$. The coupling is mediated by density 
fluctuations of local electrons which lead to the appearance of the   
charge susceptibility $C_\rho(k -k')$ of the localized 
electrons in \eqref{G58}. 

The spectral sum rule can be found by integrating  
($A_k^+(\omega)+ A_k^-(\omega)$) over $\omega$ 
and using $[c_k^\dagger,c_k]_+=1$, 
\begin{eqnarray}
                               \label{G59}
1= \int_{-\infty}^{\infty} d\omega\; (A_k^+(\omega) + 
A_k^-(\omega)) &=&
|\tilde{\alpha}_k|^2 + \frac{1}{N}\sum_{k'} 
|\tilde{\beta}_{k k'}|^2 C_\rho(k -k') . 
\end{eqnarray}
This relation can also be derived directly from \eqref{G45}.\\

Next, we evaluate the averaged one-particle  spectral functions  
$B^+(\omega)$ and  $B^-(\omega)$ of localized electrons,
\begin{eqnarray}
                      \label{G60}
B^+(\omega) &=& \frac{1}{2\pi} \int_{-\infty}^{\infty}
\big< f_i (t) \;f_{i}^\dagger
\big> \; e^{i\omega t} dt\, \hspace*{0.8cm}
B^-(\omega) = 
\frac{1}{2\pi} \int_{-\infty}^{\infty}
\big< f_{i}^\dagger  \;f_{i}(t) \big> 
\; e^{i\omega t} dt .
\end{eqnarray}
Since the local $f$-charge is a constant of motion, 
all correlation functions between different sites $i$ and $j$ vanish.  
For the evaluation we again use \eqref{G42} and the following ansatz 
for the transformed $f$-operator 
\begin{eqnarray}
                        \label{G61}
f_{i}(\lambda) &=& \nu_{\lambda} f_{i}
+   \frac{1}{N} \sum_{{k k'}} \mu_{k,k',\lambda}\; 
e^{i(k -k'){R}_i} \; \delta (c_k^\dagger 
c_{k'}) f_i  , 
\end{eqnarray}
where $\nu_{\lambda}$ and $\mu_{k,k',\lambda}$ 
are the  $\lambda$-dependent coefficients. As before, the operator structure 
of the second term is assumed to agree with the one obtained from the 
first order contribution in $U$ and higher order terms have been neglected. 
The initial parameter values are ($\lambda = \Lambda$): $ \nu_{\Lambda} = 1, 
\mu_{k k', \Lambda} =0$. The renormalization equations for 
$\nu_{\lambda}$ and $\mu_{k,k',\lambda}$ can easily be derived.  
In the limit $\lambda\to 0$  we find the following result
\begin{eqnarray}
                     \label{G65}
B^+(\omega) &=& \tilde{\nu}^2 B_0^+(\omega) 
+
\frac{1}{N^2} \sum_{k k'} |\tilde{\mu}_{k k'}|^2
f(\tilde{\varepsilon}_k) (1- f(\tilde{\varepsilon}_{k'}))
B_0^+(\omega + \tilde{\varepsilon}_k -
\tilde{\varepsilon}_{k'})\\
&& \nonumber \\
                      \label{G66}
B^-(\omega) &=&  \tilde{\nu}^2 B_0^-(\omega) 
+
\frac{1}{N^2} \sum_{k k'} |\tilde{\mu}_{k'k}|^2
f(\tilde{\varepsilon}_k) (1- f(\tilde{\varepsilon}_{k'}))
B_0^-(\omega -( \tilde{\varepsilon}_k -
\tilde{\varepsilon}_{k'}) ) .
\end{eqnarray} 
We have introduced new correlation functions
\begin{eqnarray}
                    \label{G67}
B^+_0(\omega) &=& \frac{1}{2\pi}
\int_{-\infty}^{\infty} dt e^{i\omega t}\;
\big< f_i(t) f_i^\dagger \big>_{\tilde{\cal H}_f}
\hspace*{2cm}
B^{-}_0(\omega) = \frac{1}{2\pi}
\int_{-\infty}^{\infty} dt e^{i\omega t}\;
\big< f_i^\dagger f_i(t) \big>_{\tilde{\cal H}_f}. \nonumber \\
&& 
\end{eqnarray}
which are defined with the lattice gas model $\tilde{\cal H}_f$. 
In Appendix B the functions $B_0^{\pm}(\omega)$
are evaluated by use of the retarded one-$f$ electron 
Green function ${\cal G}_0(\omega)$, defined with $\tilde{\cal H}_f$,
 \begin{eqnarray}
               \label{G68}
{\cal G}_0(\omega) &=& i  
\int_0^\infty dt e^{i(\omega +i \eta) t} \;
\big< [f_i(t), f_i^\dagger]_+ \big>_{\tilde{\cal H}_f} 
\hspace*{2cm} (\eta =0^+). 
\end{eqnarray}

\section{Results} 
\subsection{Static properties}

We now discuss the results obtained by  numerical evaluation 
of the renormalization equations for the one-dimensional lattice. Note
  that in the following all energy values are given in 
units of the hopping matrix element $t$. One first has to 
calculate the fully renormalized parameters 
$\tilde{\varepsilon}_k$, $\tilde{\varepsilon}_f$, and
$\tilde{g}_{ij}$ which determine the effective Hamiltonian   
$\tilde{\cal H}= \tilde{\cal H}_c + \tilde{\cal H}_f + \tilde{E}$
of \eqref{G41}.
All static expectation values which enter the renormalization equations 
are evaluated self-consistently with the full Hamiltonian, 
following the procedure  
outlined in  Sections III and IV.  
The effective Hamiltonian $ \tilde{\cal H}_{f} $ describes the 
lattice gas model of ions which interact via a possibly long range 
interaction $\tilde{g}_{ij} $. 
Thus, we use an additional  Monte Carlo simulation to evaluate the expectation values
$n^{f} = \left\langle n_{i} \right\rangle_{\tilde{\cal H}_{f}} $ and 
$C_{ff}(\kappa)= \left\langle \delta n_{i} \delta n_{j} \right\rangle = 
\left\langle \delta n_{i} \delta n_{j}\right\rangle_{\tilde{\cal H}_{f}}$, 
which enter the renormalization equations  together with 
$\left\langle c_k^{\dag} c_k \right \rangle_{\tilde{\cal H}_{c}} 
=[1+ \displaystyle e^{\beta \tilde{\varepsilon}_k }]^{-1}$.

As already mentioned, we start by making a guess for the initial values of $n_f $,
$C_\rho(k- k')$ and $\big< c_k^\dagger c_k  \big>$, 
and calculate the renormalized parameters for $\lambda \rightarrow 0$. 
That is,  we find $\tilde{\varepsilon}_f=\varepsilon_{f,\lambda=0}$,
$\tilde{\varepsilon}_k= \varepsilon_{k, \lambda=0}$, 
and $\tilde{g}_{q}= g_{{q}, \lambda=0}$ 
by solving numerically the renormalization equations \eqref{G33}--\eqref{G39}. 
Next, using the effective coupling $\tilde{g}_{q}$ we find the new values 
of the spatial interactions $\tilde{g}_{ij}$ and then recalculate 
the localized electron expectation values $n_f$ and $C_{ff}(\kappa)$ 
using a classical Monte Carlo simulation. 
The Fourier transform of $C_{ff}(\kappa)$ gives an approximate 
expression for $C_\rho(k-k')$.  
Using new  values of  $n_f $, $C_\rho(k-k')$ and 
$\big< c_k^\dagger c_k \big>$ 
we iterate  \eqref{G33} -- \eqref{G39}  until a self-consistent solution is reached. 

In the following we consider only the one-dimensional 
Falicov-Kimball model at half-filling, 
\begin{equation}
\label{G75}
n_{f} + n_{c} = 1 , 
\end{equation}
where  
$n_f = \langle n_i \rangle$ and $n_c= 1/N \sum_k\langle 
c_k^\dagger c_k\rangle $ 
is the averaged local occupation numbers of conduction electrons.
Condition \eqref{G75} represents a point of special interest for
valence and possibly metal-insulator transitions, caused by the promotion
of electrons from localized $f$ orbitals  ( $f^n \rightarrow f^{n+1}$ )
to the conduction band states\cite{Ott}. 


\subsubsection{Temperature dependence}

\begin{figure}
\begin{center}
\includegraphics[width=8cm,height=6cm]{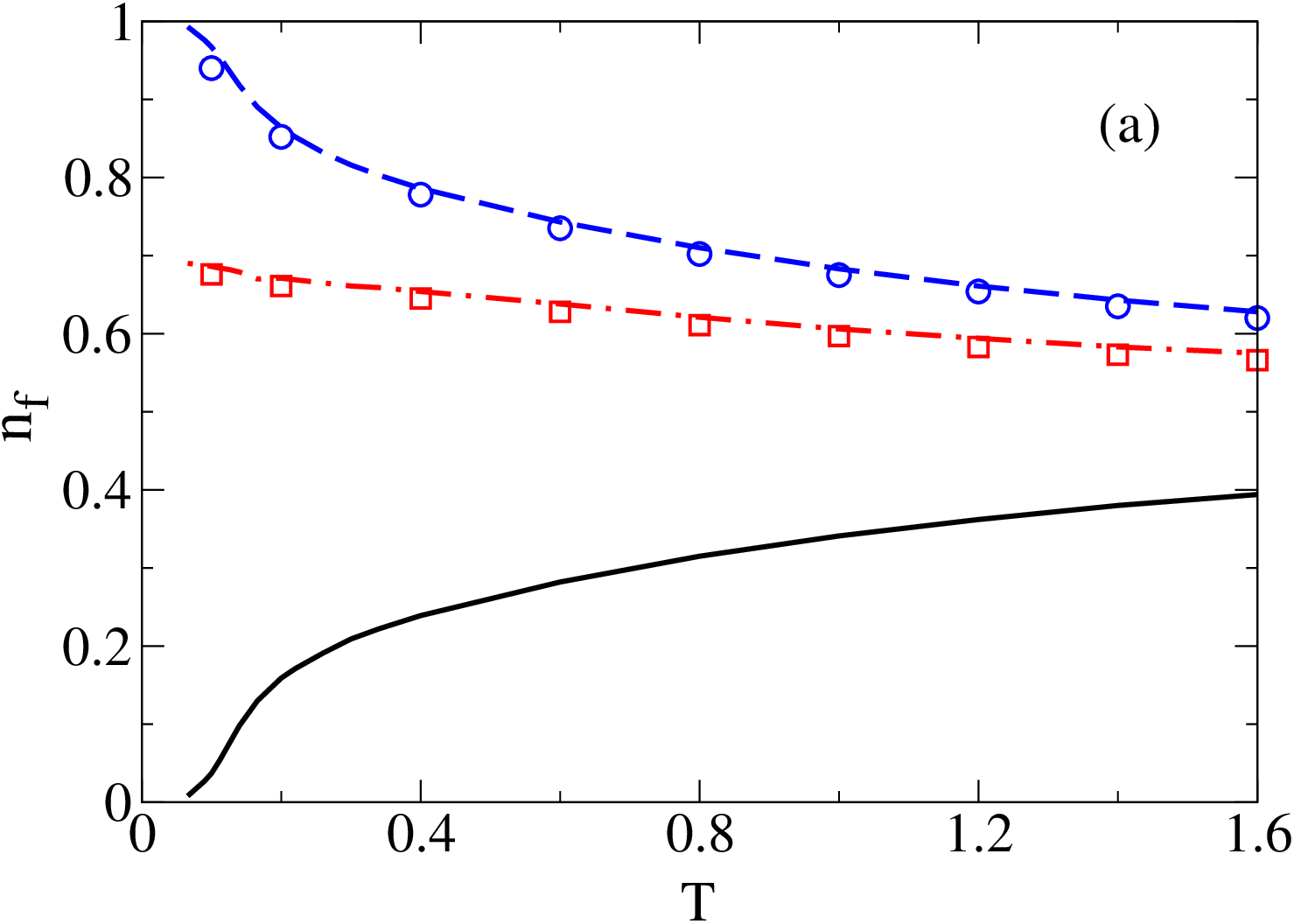}
\includegraphics[width=8cm,height=6cm]{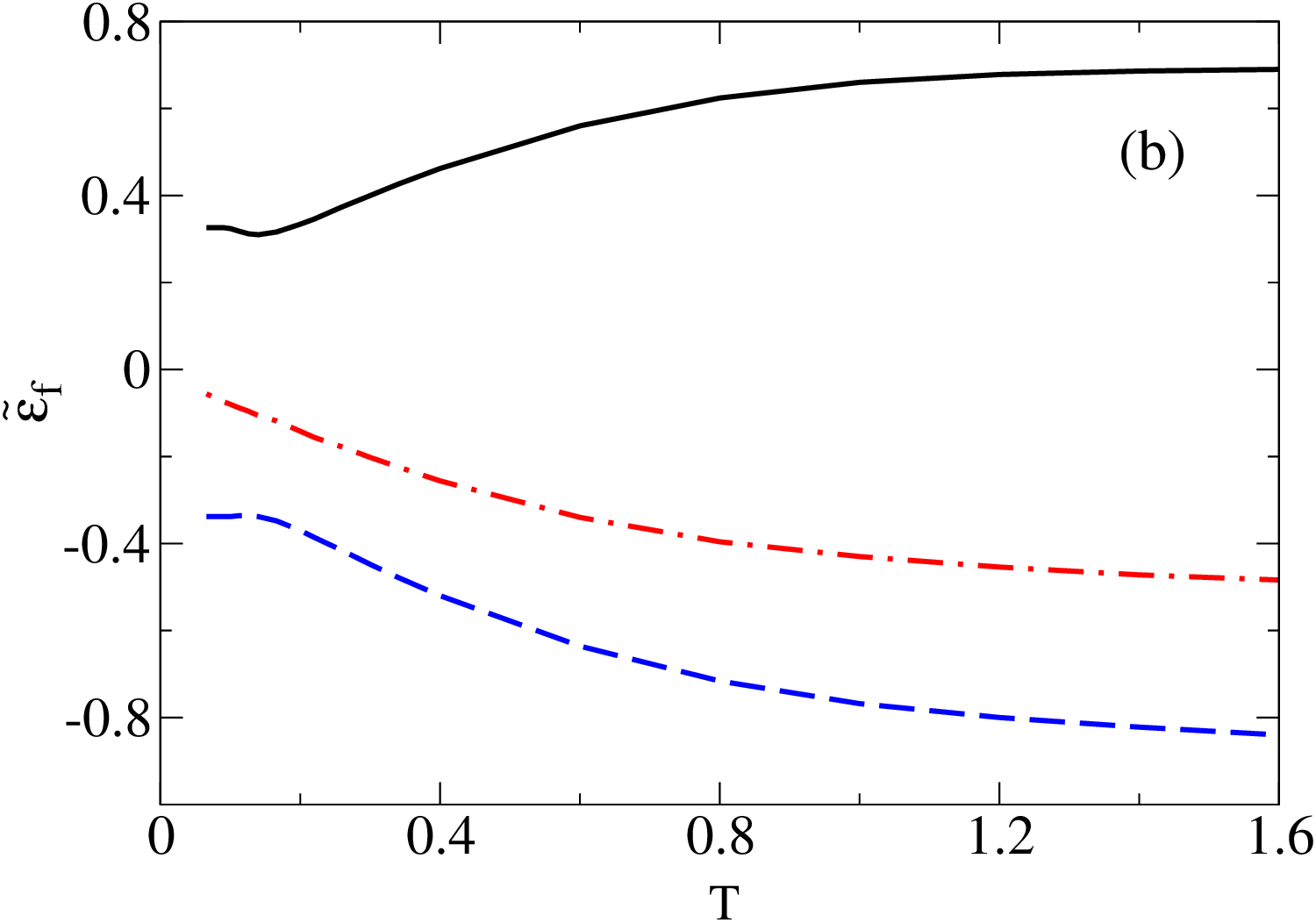}
\end{center}
\caption{(a) $f$-electron occupation number $n_f$
as function of temperature for  $U=1.0$ and for three different 
values of the bare $f$-level $\varepsilon_{f} = 0.5$ (dashed), 
$\varepsilon_{f} = 1.2$ (dashed dotted), and $\varepsilon_f=3.5$ (solid). 
The open circles ($\varepsilon_{f} = 0.5$)
and open squares ($\varepsilon_{f} = 1.2$)
refer to ED results of reference \cite{QMC}. Note the 
good agreement with the PRM results. The values of $\varepsilon_{f}$ 
are measured from the lower edge of the conduction band. (b) Renormalized 
$f$-electron energy $\tilde{\varepsilon}_f$ 
as function of temperature for the same values of  $U$ and 
$\varepsilon_{f}$-values as in panel (a). }
\label{fig:1}
\end{figure}

In Fig.~\ref{fig:1} the averaged local $f$-occupation number 
$n_f$ and the renormalized 
$f$-energy $\tilde{\varepsilon}_f$ from the PRM approach 
are shown as function of temperature for $U=1.0$ 
and several values of unrenormalized 
{\it f}-level position $\varepsilon_f$. 
Note, the chemical potential $\mu$ has to be adjusted 
at each temperature $T$, so as to fulfill condition \eqref{G75}. 
In the evaluation procedure this requires extra numerical cycles, 
in addition to those from the  self-consistent determination of the expectation values. 
Therefore, only a rather small system of 20 lattice sites has been used 
to generate Fig.~\ref{fig:1}. 
Also the considerably 
increased effort in the  MC sampling 
for the lattice gas part in ${\cal H}_{0,\lambda}$  
at higher temperatures suggest to restrict oneself to a rather 
small system. This will be discussed below in some more detail.  
As expected, for small values of the bare $f$-level $\varepsilon_f$ 
(dashed curve in Fig.~\ref{fig:1}a),
the average $f$-occupation $n_f$ increases with decreasing temperature
and tends to $1$ at $T=0$.  The corresponding values of the renormalized 
$f$-level remain below the chemical potential ($\tilde{\varepsilon}_f <0$
in Fig.~\ref{fig:1}b), so that all the sites are occupied with one 
$f$-electron. 
In contrast, for large values of  $\varepsilon_f$ (solid curves) $n_f$ tends 
to zero for $T \rightarrow 0$ and $\tilde{\varepsilon}_f$ stays above the 
chemical potential ($\tilde{\varepsilon}_f >0$). Thus all the $f$-sites are 
empty and the renormalized conduction electron band is completely filled. 
Finally, at intermediate values of  the bare $\varepsilon_f$-value (dashed
dotted  curves) 
the $f$ occupation at $T=0$ tends to some value between $0$ and $1$,  
signalizing an intermediate $f$-valence. 
In that case, the renormalized $f$-level approaches the chemical potential 
($\tilde{\varepsilon}_f =0$) at $T=0$. Below the chemical potential
all the conduction electron states are occupied and $f$-states only partially, 
due to the condition $n_f + n_c=1$. 
 
\subsubsection{Low temperature properties at fillings $n_{f} = 1/2$ and 
$n_{f} = 1/3$}
Next, we discuss static properties for $n_f= 1/2$ and $1/3$ at low temperatures. 
The parameters $\mu$ and $\varepsilon_{f}$ have to be adjusted in such a
way that the condition  \eqref{G75} is fulfilled for a given value of $n_f$. 
Note that in a Monte Carlo simulation 
a nonzero temperature has to be used, which is set equal to
$k_{B} T= 0.1$.  Also the most dominant
ground-state configurations have been incorporated
in the Monte-Carlo simulation 
so that the system size could be extended to $N = 150$ sites.
For $n_{f} = 1/2$ the chemical potential $\mu$ is in the middle of 
the renormalized conduction band $\tilde{\varepsilon}_k$.

\begin{figure}[h]
\begin{center}
\includegraphics[width=8cm,height=6cm]{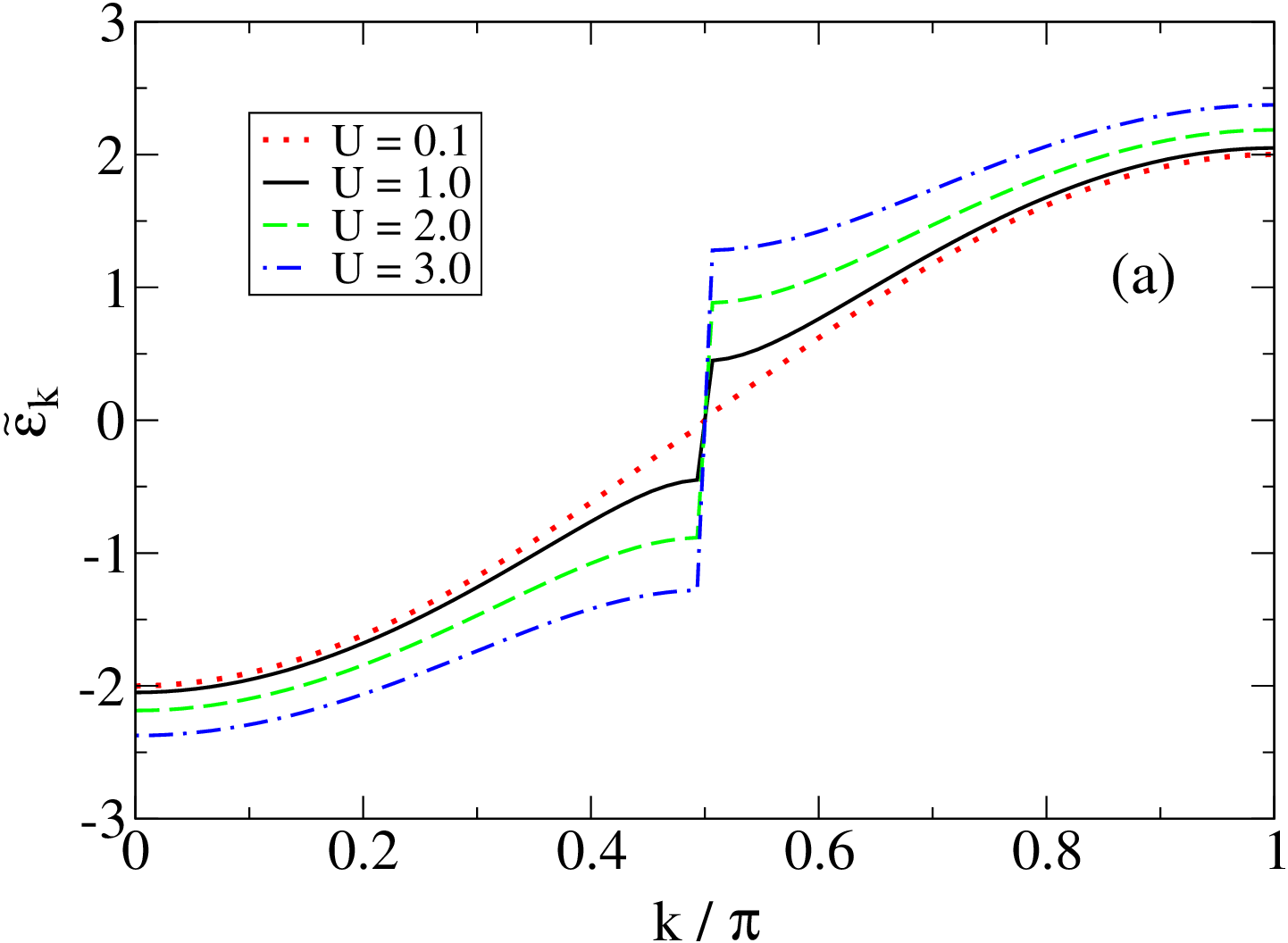} \vspace*{0.9cm} \hspace*{1mm}
\includegraphics[width=8cm,height=6cm]{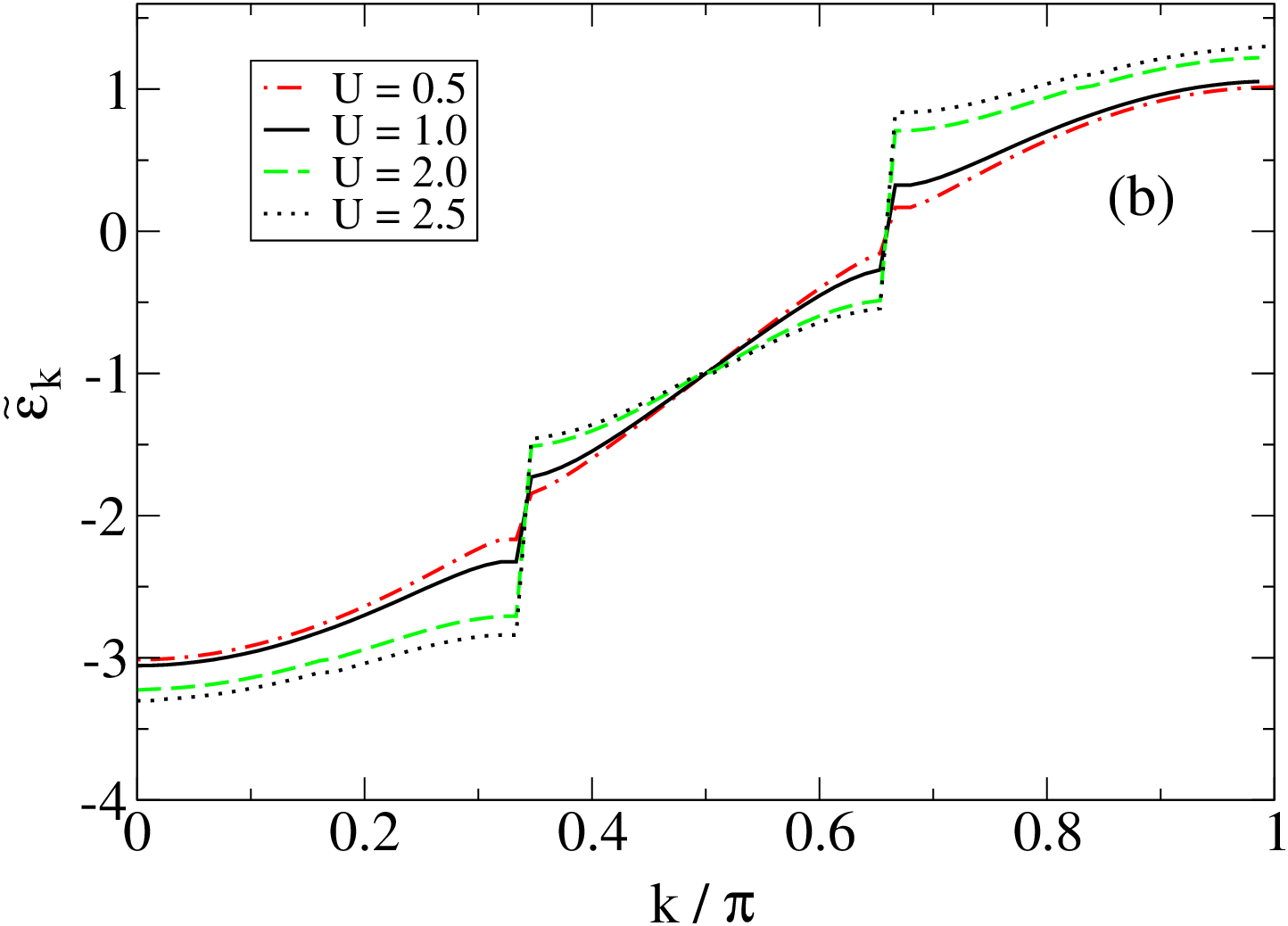} 
\includegraphics[width=8cm,height=6cm]{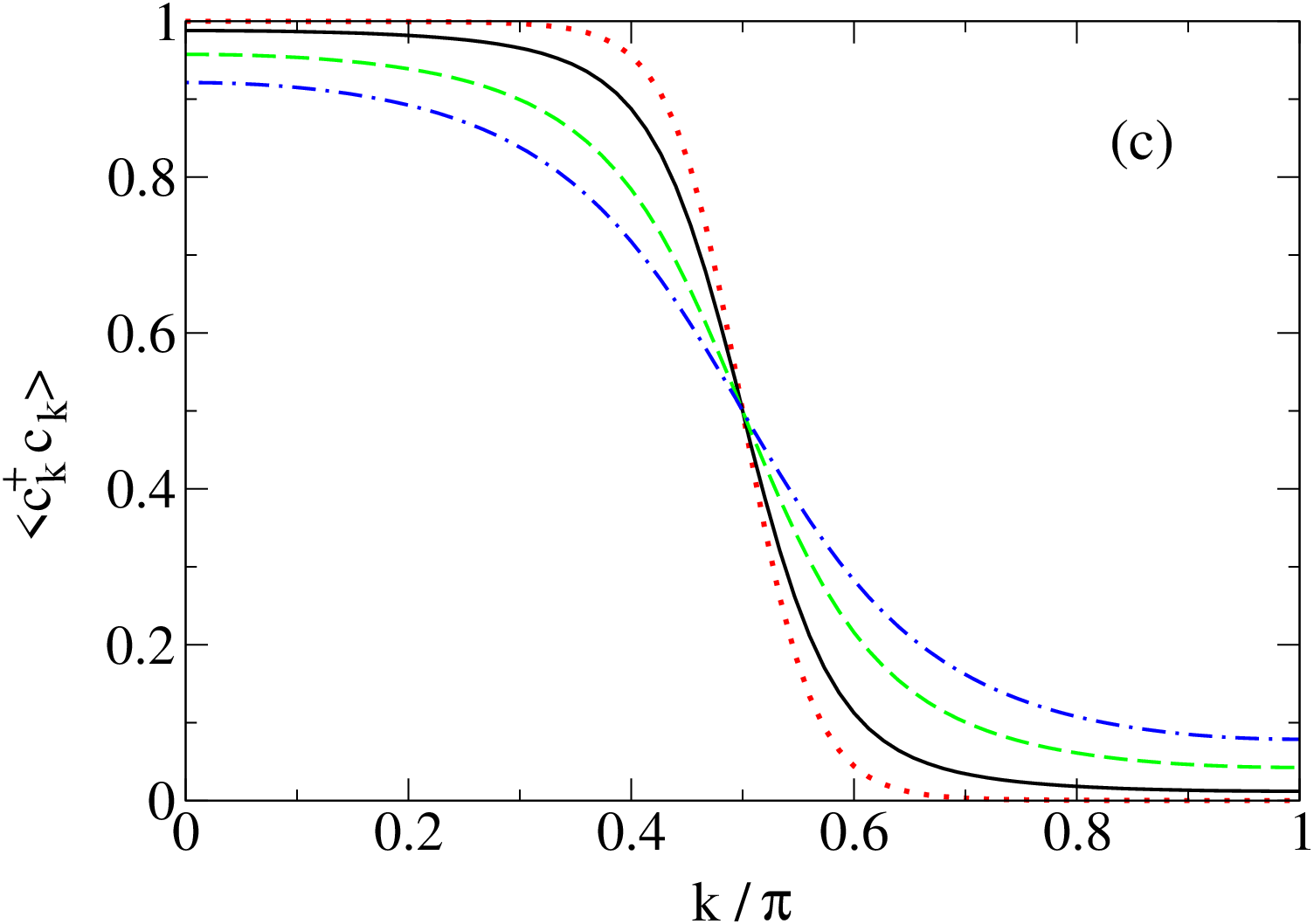} \hspace*{1mm}
\includegraphics[width=8cm,height=6cm]{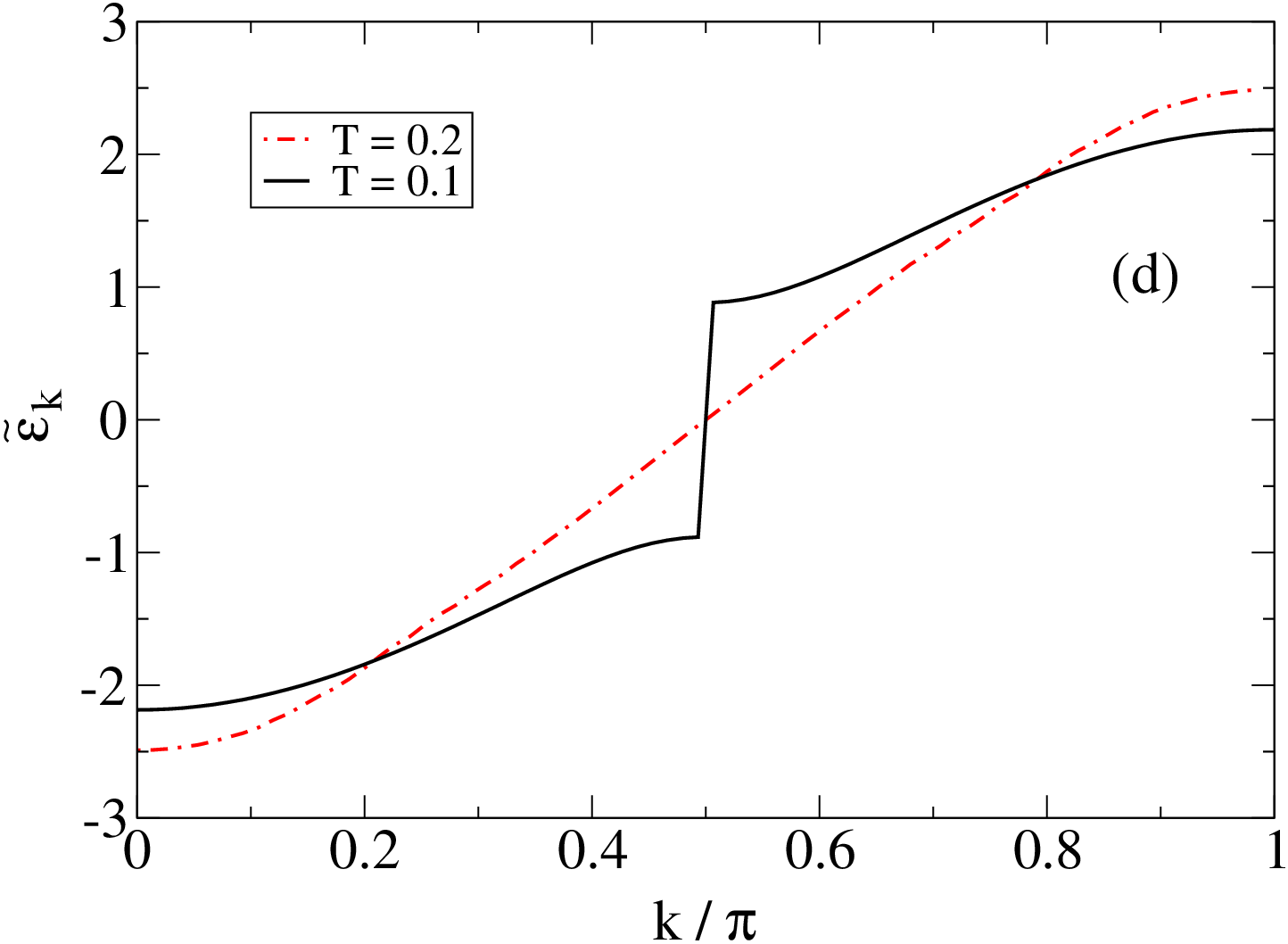}
\end{center}
\caption{(a) Renormalized conduction electron energy 
$\tilde{\varepsilon}_k$ for different Coulomb  
repulsions $U$ ($= 0.1,1.0,2.0,3.0$) and $n_{c} = n_{f}= 1/2$. 
Note the appearance of  a gap of magnitude $U$ at the Fermi momentum $k_F
= \pi/2$. 
(b) Same quantity 
$\tilde{\varepsilon}_k$ for $n_{c} = 2/3$ ($n_f= 1/3$) for 
$U= 0.5, 1.0, 2.0, 2.5$.  $\tilde{\varepsilon}_k$
shows two gaps where the one at the 
Fermi momentum $k_F=(2/3) \pi$ is now smaller than $U$.(c) Momentum
distribution $n_k= \langle c_{k}^{\dag} c_k\rangle $ for the same
    parameters as in panel (a) and $n_{f} = n_{c}= 1/2$. 
    The smooth behavior of $n_k$ 
    at the Fermi momentum follows from the gap in 
    $\tilde{\varepsilon}_k$. 
    (d) $\tilde{\varepsilon}_k$ for two different 
    temperatures $T=0.1$ (solid line) and 
    $T=0.2$ (dashed dotted line) and $U=2.0$, $n_f=1/2$. 
    Note that the gap at $k_F$ for $T=0.1$ 
    is completely smeared out for the higher temperature $T=0.2$.} 
\label{fig:5}
\end{figure}
  

In Fig.~\ref{fig:5}c the result for  $n_k$ obtained from the PRM
 is plotted as function of $k$ for $n_c=1/2$  
and for different values of the Coulomb repulsion $U$. 
The results for $n_k$ from the PRM approach are very similar to
those from Ref.~\cite{farkasovsky.03} (for the same
values of $U$).

In Fig.~\ref{fig:5}a  
the renormalized one-particle energy 
$\tilde{\varepsilon}_k$  is shown as function of momentum $k$ 
for $n_c=1/2$ ($n_f =1/2$), and 
for the same interaction values $U$ as in Fig.~\ref{fig:5}c.
Note, that $\tilde{\varepsilon}_k$ can be considered as 
the excitation energy of a quasiparticle of the original interacting model 
\eqref{G1}. This follows from the diagonal form of 
$\tilde{\cal H}_c$ in \eqref{G41} and the fact that $\tilde{\cal H}$ 
is obtained from the original model by a unitary transformation. 
The main feature of $\tilde{\varepsilon}_k$ 
is the appearance of 
a gap at the Fermi level, which opens for any, even small, value 
of the repulsion energy $U$. Such a gap was recently anticipated by 
Farkasovsky \cite{farkasovsky.03} from the correspondence 
of the Falicov-Kimball model to  
the Tomonaga-Luttinger fermions. There, the smooth behavior of  
$n_k$  with $k$ also gives rise to a
gap in the charge excitations.    
Note that for $n_c=2/3$ ($n_f=1/3$) (Fig.~\ref{fig:5}b)
again a gap opens at the Fermi 
energy for $k_F = (2/3) \pi$ for all values of $U$. Thus, the
Luttinger theorem is fulfilled even for the present case of an insulating
state. By closer inspection of Figs.~\ref{fig:5}a,b one finds that the gaps 
at $k_F$ scale linearly with $U$ both for $n_c=1/2$ and $n_c=2/3$. The
proportionality constant is almost $1$ for $n_c=1/2$ whereas for $n_c=2/3$ 
it is somewhat smaller than $1$. \vspace{0.5cm}

\begin{figure}[h]
\begin{center}
\includegraphics[width=8cm,height=6cm]{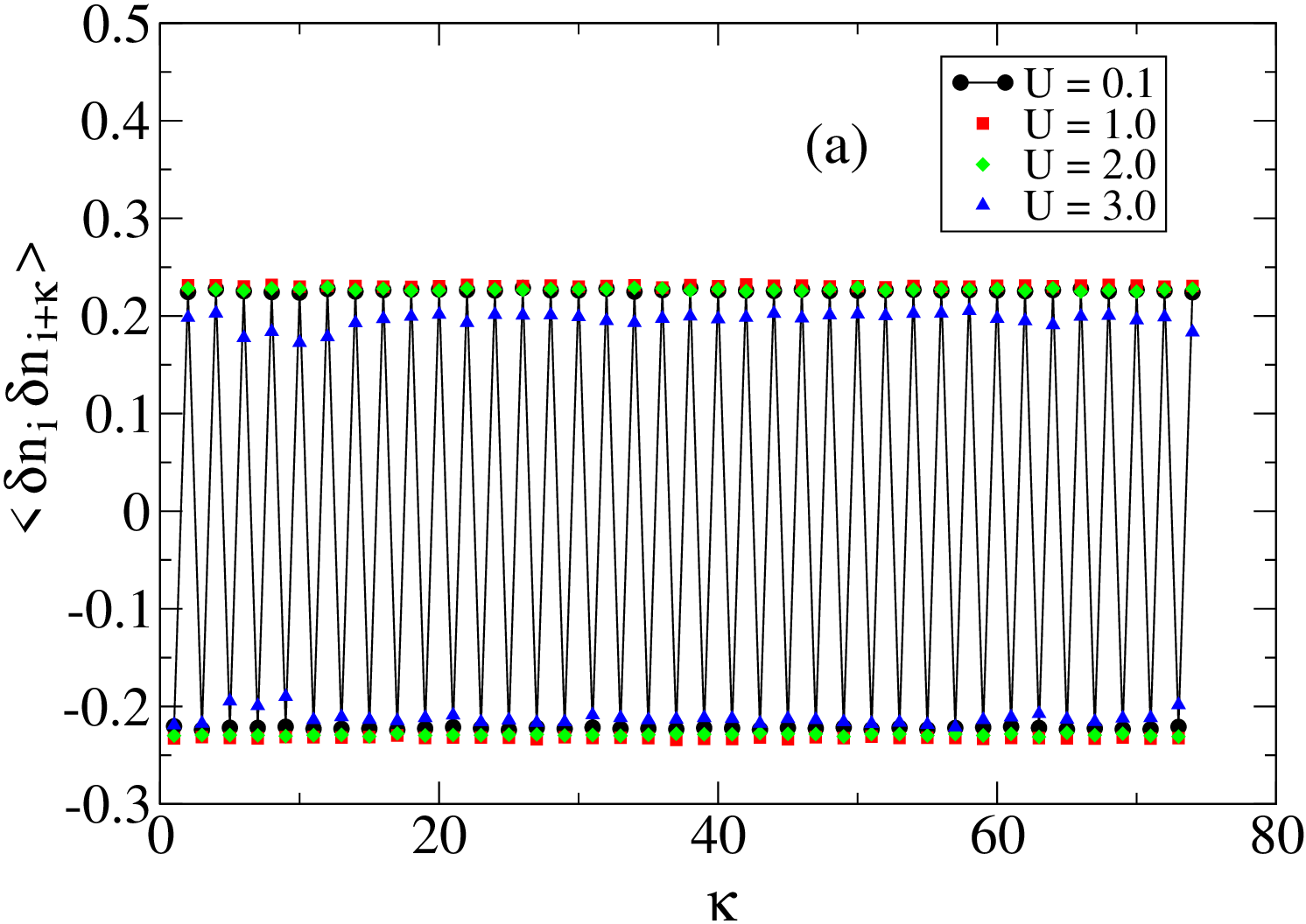}
\includegraphics[width=8cm,height=6cm]{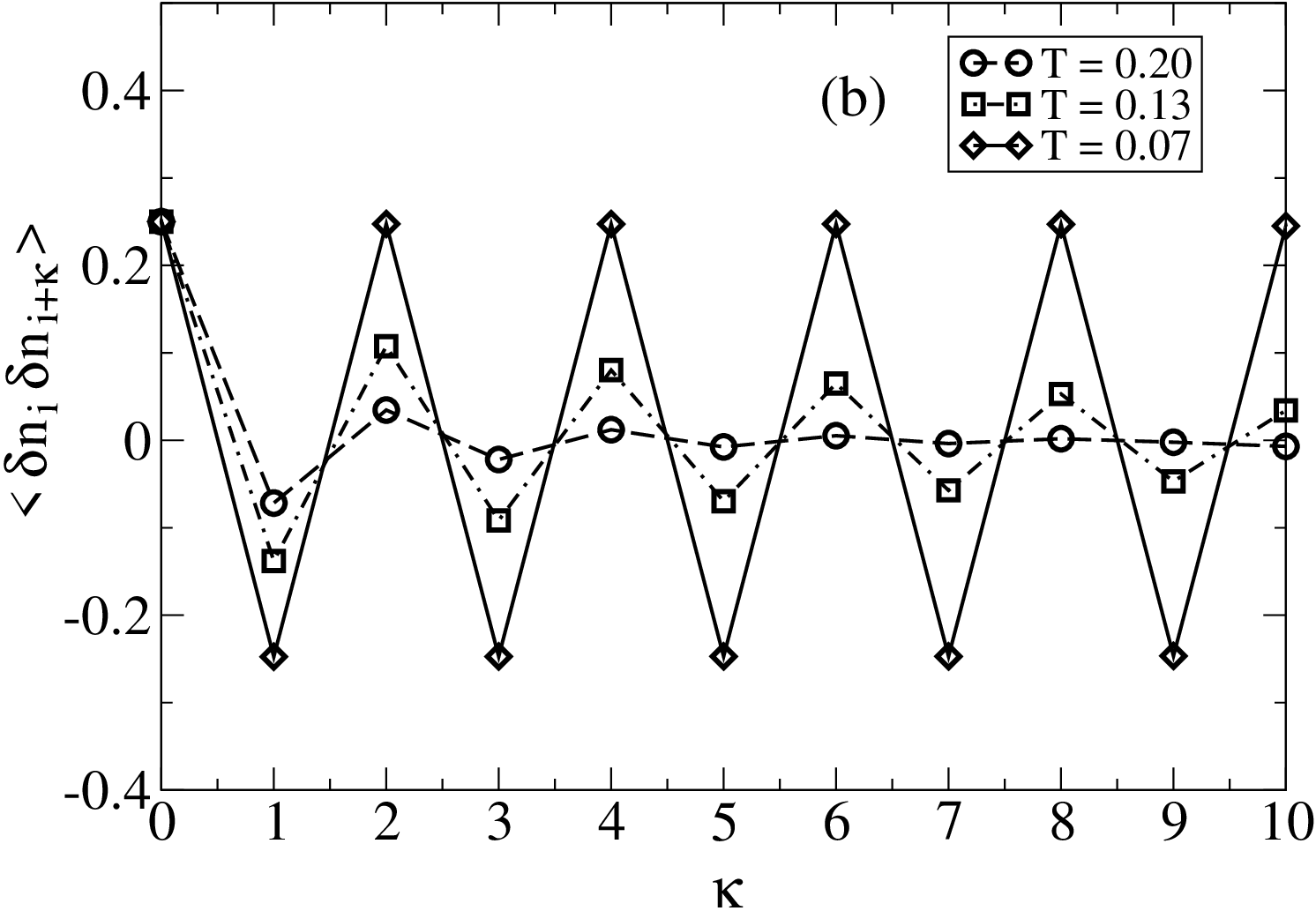}
\end{center}
\caption{(a) Correlation function $C_{ff}(\kappa)= \langle \delta n_{i}
  \delta n_{i+\kappa} \rangle$ of localized electrons for 
$n_f =1/2$. For $n_f =1/3$ we obtain an analogous result. The length 
 $\kappa$ characterizes the distance between the sites. The results 
correspond to averages $(1/N)\sum_i \langle \delta n_{i}
  \delta n_{i+\kappa} \rangle$ over all lattice sites (see text). (b) 
Correlation function $C_{ff}(\kappa) =\langle \delta n_{i}
  \delta n_{i+\kappa} \rangle$ of localized electrons for a chain with
  20 lattice sites for three different temperatures $T=0.20, 
0.13, 0.07$ and $U = 1.0$ and
  $n_{f} = 1/2$.}
\label{fig:7}
\end{figure}

Fig.~\ref{fig:7}a shows the spatial correlation functions 
$C_{ff}(\kappa)= \langle \delta n_{i}
  \delta n_{i+\kappa} \rangle$ between the localized electrons for the same
  parameters as in Fig.~\ref{fig:5}a. 
The index $\kappa$ denotes the distance
between the two involved lattice sites. Note that for  
$n_f=1/2$ the values of the correlation function
are always the same at every second site ($\kappa= 2,4,6, \cdots$) whereas 
for all sites in between ($\kappa= 1, 3, 5, \cdots $) they differ only by a 
minus sign. 

\subsubsection{Discussion}

Note that the long-range correlations develop  
already for arbitrarily small values of $U$. This behavior should 
be related to the long range order of the
one-dimensional Falicov-Kimball model at zero temperature~\cite{kennedy.86}. 
Note that in the present calculations the 
temperature is chosen small but 
finite ($k_BT= 0.1$),  so that in the strict sense no infinitely long 
range order exists. The origin of the long-range correlations 
can easily be understood from a low
order $U$ expansion of the effective $f$-coupling. 
For the Fourier transform  $g_{q}$ of the local quantity  
$g_{ij}$ one finds by use of \eqref{G39}, \eqref{G28} and \eqref{G30} 
\begin{eqnarray}
\label{G75a}
g_{{q}, (\lambda - \Delta \lambda)} &=&   g_{{q}, \lambda}
- \frac{U^2}{2N} \sum_k 
\frac{\big< c_k^\dagger c_k\big> - 
\big<c_{k+{q}}^\dagger c_{k+{q}}\big>}
{\varepsilon_{k+{q},\lambda} - \varepsilon_{k, 
\lambda}} \;
\Theta_{k, k+{q}}^{\lambda, \Delta \lambda} \\
&& \hspace* {1cm} +
 \frac{U^2}{2N^2} \sum_{k, {q'}} 
\frac{\big< c_k^\dagger c_k\big> - 
\big<c_{k+{q'}}^\dagger c_{k+{q'}}\big>}
{\varepsilon_{k+{q'},\lambda} - 
\varepsilon_{k, \lambda}} \;
\Theta_{k, k+{q'}}^{\lambda, \Delta \lambda} \nonumber
\end{eqnarray}
where the last term excludes equal sites $i=j$ in $g_{ij}$.  
As follows from \eqref{G75a}, the interaction between the localized electrons
is caused from the coupling to particle-hole excitations 
${\varepsilon_{k+{q},\lambda} - \varepsilon_{k, 
\lambda}}$ where the wavevector runs over the whole Brillouin zone. Here,
the energies $\varepsilon_{k,\lambda}$ and  
$\varepsilon_{k+{q},\lambda}$ have to be either below or above the Fermi 
level, i.e. $|k|< k_F$  and $|k+q|> k_F$ or  $|k|>k_F$  and $|k+q|< k_F$.
Note that for $q=\pi$ and $n_f=1/2$
this condition is fulfilled for all $k$-values
of the Brillouin zone so that the most dominant renormalization of 
$g_{{q}}$ occurs for $q=\pi$.    
In contrast, for ${q}$-values different from $\pi$ always  less
$k$ terms  contribute.
For instance, for $q\approx 0$ only $k$ points from the sum in 
\eqref{G75a} can contribute which are located in a small 
region around the Fermi momentum $k_F$. As a result of  
the dominant coupling $g_{q}$ at $q=\pi$ one is immediately lead 
to a infinitely ranged spatial correlation between the 
$f$-electrons having an alternating sign from site to site. 
This explains the long-range behavior in Fig.~\ref{fig:7}a. 
Note again that this feature is not based
on a strong-coupling argument. Instead, 
the long-ranged correlations are observed already for 
small values for $U$.   \\

The alternating behavior of $C_{ff}(\kappa)= \big< \delta n_i
\delta n_{i+\kappa} \big>$ with $\kappa$ can be interpreted as follows:  
At zero temperatures the model has two equivalent
degenerate ground states of period two (for $n_f=1/2$), 
separated by a potential barrier, with the arrays $\{1010 . . . 10\}$ and 
$\{0101 ... 01\}$ of the localized ions. Therefore, we might use
the following ansatz for the state vector of the system
at very low but finite temperatures,       
\begin{eqnarray}
\label{G76}
|\Phi_g^{(2)}\rangle = \frac{1}{\sqrt{2}} \big(
|10101 ... 10 \rangle + |0101 ... 01 \rangle
\big) \ .
\end{eqnarray}
In order to obey translation invariance the sum of the 
two generated ground states have been taken and 
ground-state fluctuations have been neglected. Note that the 
two states are never  connected by nonvanishing matrix elements when physical 
quantities are evaluated. 
To proceed, we use \eqref{G76} to evaluate the correlation function 
$\langle n_i n_{i+\kappa} \rangle$ of Fig.~\ref{fig:7}. As explained,
the result corresponds to the average $(1/N)\sum_i 
\langle n_i n_{i+\kappa} \rangle$ over all lattice sites and is independent of
$i$. From ansatz \eqref{G76} one finds
\begin{equation}
\label{G77}
\langle n_i n_{i+\kappa} \rangle = \left\{ 
\begin{array}{lll}
\frac{1}{2}   &  \mbox{:} &    \kappa= 2, 4, 6, ... \\
 0            & \mbox{:}  &   \kappa= 1, 3, 5, ... 
\end{array} 
\right.
\end{equation}
With $\big< n_i \big> =1/2$ one obtains for 
$C_{ff}(\kappa) = \langle \delta n_i \; 
\delta n_{i+\kappa} \rangle$  
\begin{equation}
\label{G78}
C_{ff}(\kappa) = \left\{ 
\begin{array}{rll} 
\frac{1}{4} &  \mbox{:}  & \kappa= 2, 4, 6, ... \\
- \frac{1}{4}  & \mbox{:} &  \kappa= 1, 3, 5, ... 
\end{array} 
\right.
\end{equation}
Obviously  the result \eqref{G78} agrees very well with the 
outcome from the PRM approach according to Fig.~\ref{fig:7}a 
for small but also for rather large values of $U$.  A similar reasoning may 
also be given for $n_f =1/3$ ($n_c=2/3$). 
Here, the degenerate ground-states have period three with arrays $\{100100 ... 100 \}$, 
$\{010010 ... 010 \}$, and $\{001001 ... 001 \}$ in analogy to \eqref{G76}.

Note that the amplitude of the $f$-correlation function 
$\langle \delta n_i \; \delta n_{i+\kappa} \rangle$ is almost independent of
$U$. Already for small $U$-values the 
$f$-electrons are ordered according to the qualitative description
of Fig.~\ref{fig:7}a. The behavior of the conduction 
electrons is quite different. 
The correlation function $C_{cf}(\kappa)= 
\big < n_i^c n_{i+\kappa}\big>$ describes the spatial correlation 
between a conduction electron at site $i$ and an $f$ electron at 
site $i+\kappa$. An evaluation for $C_{cf}(\kappa)$ 
correct up to first order in $U$ gives
\begin{eqnarray}
\label{G78a}
C_{cf}(\kappa) &=& \frac{n_f}{N} \sum_{k}
|\tilde{\alpha}_k|^2 \big< c_k^\dagger c_k \big>_{\tilde{\cal H}_c}
+
\frac{2}{N} (-1)^\kappa \sum_k \tilde{\alpha}_k
\tilde{\beta}_{k+ {Q}}
\big< c_k^\dagger c_k \big>_{\tilde{\cal H}_c}
+ O(U^2)
\end{eqnarray}
$(Q=\pi)$, where the renormalized parameter $\tilde{\beta}_{k+ {Q}}
\sim U$ is
negative (compare \eqref{G47}). Obviously, the second term in \eqref{G78a}
describes an oscillating behavior with $\kappa$ of 
$cf$-correlations which favour the presence of conduction 
electrons in between the $f$ sites. However, 
in contrast to the oscillating behavior of $C_{ff}(\kappa)$, 
the amplitude of this charge-density 
wave-like behavior of the correlation function is proportional to $U$. \\

For an interpretation of the temperature dependence of 
$C_{ff}(\kappa)$ note that the two 
arrays $\{1010 . . . 10\}$ and 
$\{0101 ... 01\}$ (for $n_f=1/2$) of the localized ions 
are equivalent to the two N\'eel states $|\uparrow \downarrow 
\uparrow \downarrow \cdots \uparrow \downarrow \rangle$ and 
$|\downarrow \uparrow \downarrow \uparrow \cdots 
\downarrow \uparrow \rangle$
of the one-dimensional Ising antiferromagnet. For the latter model
the spatial spin correlation function  $\langle S_i^z S_{i+\kappa}^z\rangle$ 
decays exponentially at low temperatures 
as $\exp{(-\kappa/\xi)}$ due to excitation of kinks or
domain walls. The correlation length is given by  
$ \xi \sim \exp{(\beta |J_z|/2)}$, which shows that it becomes infinite 
only exactly at zero temperature. $J_z/2$ is the characteristic 
energy for the creation of domain walls \cite{nagaosa.98}. 
For the one-dimensional Falicov-Kimball model a quite similar behavior is
expected for the correlation function $C_{ff}(\kappa)$. Here,  
also domain wall states can be thermally excited. 
They are obtained by interchanging the positions of neighboring $f$- and 
conduction electrons. In order to investigate the  
temperature dependence of the correlation length   
we have evaluated  
$C_{ff}(\kappa)= \langle \delta n_{i} \delta n_{i+\kappa} \rangle$ 
for different temperatures $T$ (Fig.~\ref{fig:7}b).
Note that for higher temperatures excited states would have to be included
in the MC-sampling which leads to considerable numerical increase. 
Therefore, instead of taking a system with 150 lattice sites as before
a rather small system with only 20 sites was investigated where an 
exact diagonalization was performed.   
Fig.~\ref{fig:7}b shows that  
for the two higher temperatures $T=0.20$ and $T=0.13$
the correlation function decays roughly exponentially
to zero within a few lattice sites whereas for the lowest 
temperature $T=0.07$ no decay is observed. Thus, in the 
latter case the correlation length $\xi$ obviously  exceeds the length of 
the lattice. 
The same feature was already found in Fig.~\ref{fig:7}a, where no
decay was observed up to $\kappa_{max} \approx 70$. 
We can conclude that for the lowest temperature 
no excited states  
contribute to the correlation function for the given lattice sizes.  
Note that this result is not a consequence of approximations 
used in the present PRM approach. Instead at the lowest temperature 
the correlation length is larger than the 
lattice extension but should be still finite. 
Only exactly at $T=0$ the correlation 
length $\xi$ should be infinite.  An equivalent behavior is also found for the 
itinerant electron-dispersion $\tilde{\varepsilon}_k$ as function of 
temperature. 
In Fig.~\ref{fig:5}d  the dispersion of $\tilde{\varepsilon}_k$ 
is shown for two 
temperatures $T=0.1$ and $T=0.2$ for $U=2.0$ and $n_f = 1/2$. 
Whereas at $T=0.1$ the gap at $k_F$ is clearly seen it is
interesting to note that it is already smeared out for $T=0.2$.  \\


Note that an infinitely large correlation length $\xi$ leads for the 
Fourier transform $C_\rho({q})$ of $C_{ff}(\kappa)$ to 
$C_\rho({q}) \sim  N \delta_{q, \pi}$.
Instead, for a finite correlation length $\xi$ one would find a 'softened' 
Kronneker function around ${q}=\pi$ within a range of the  
order of the inverse correlation length $\xi^{-1}$. 
Thus the gap in the one-particle 
excitation energy $\tilde{\varepsilon}_k$ of Figs.~\ref{fig:5}a,b
also softens over a range in $k$-space of the same order $\xi^{-1}$.
It folllows that the system is metallic at arbitrarily small 
temperature. The softening of the gap could hardly be seen
in Figs.~\ref{fig:5}a,b for a correlation length larger than the lattice
extension of, may be, 100 times the lattice constant. 
Thus, the results for   $\tilde{\varepsilon}_k$ and other quantities
are extremely good approximations also for the $T=0$ limit. \\


\subsection{Dynamical properties at low $T$}

\begin{figure}
\begin{center}
\includegraphics[width=5.4cm,height=4.2cm]{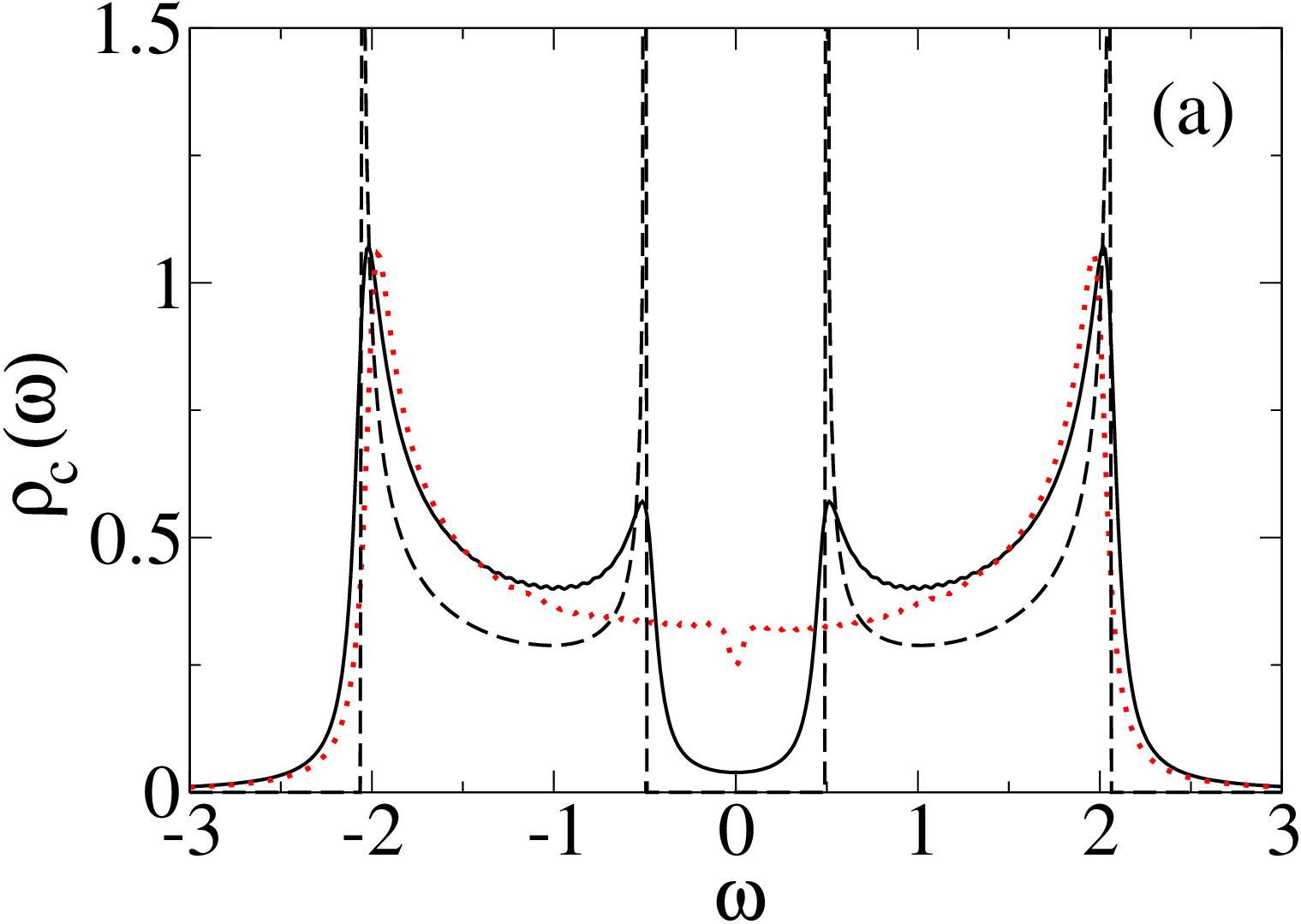}
\includegraphics[width=5.4cm,height=4.2cm]{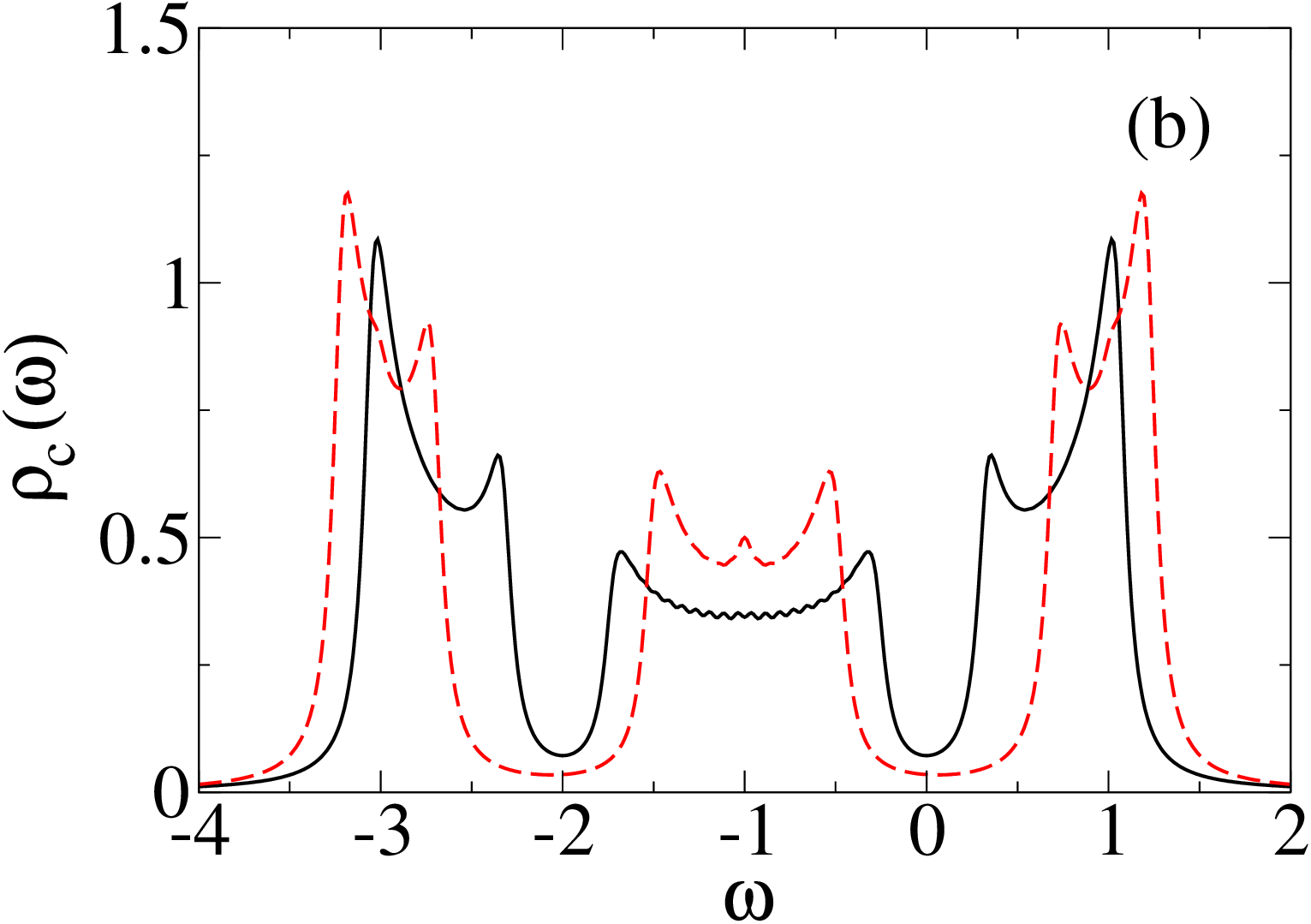}
\includegraphics[width=5.4cm,height=4.2cm]{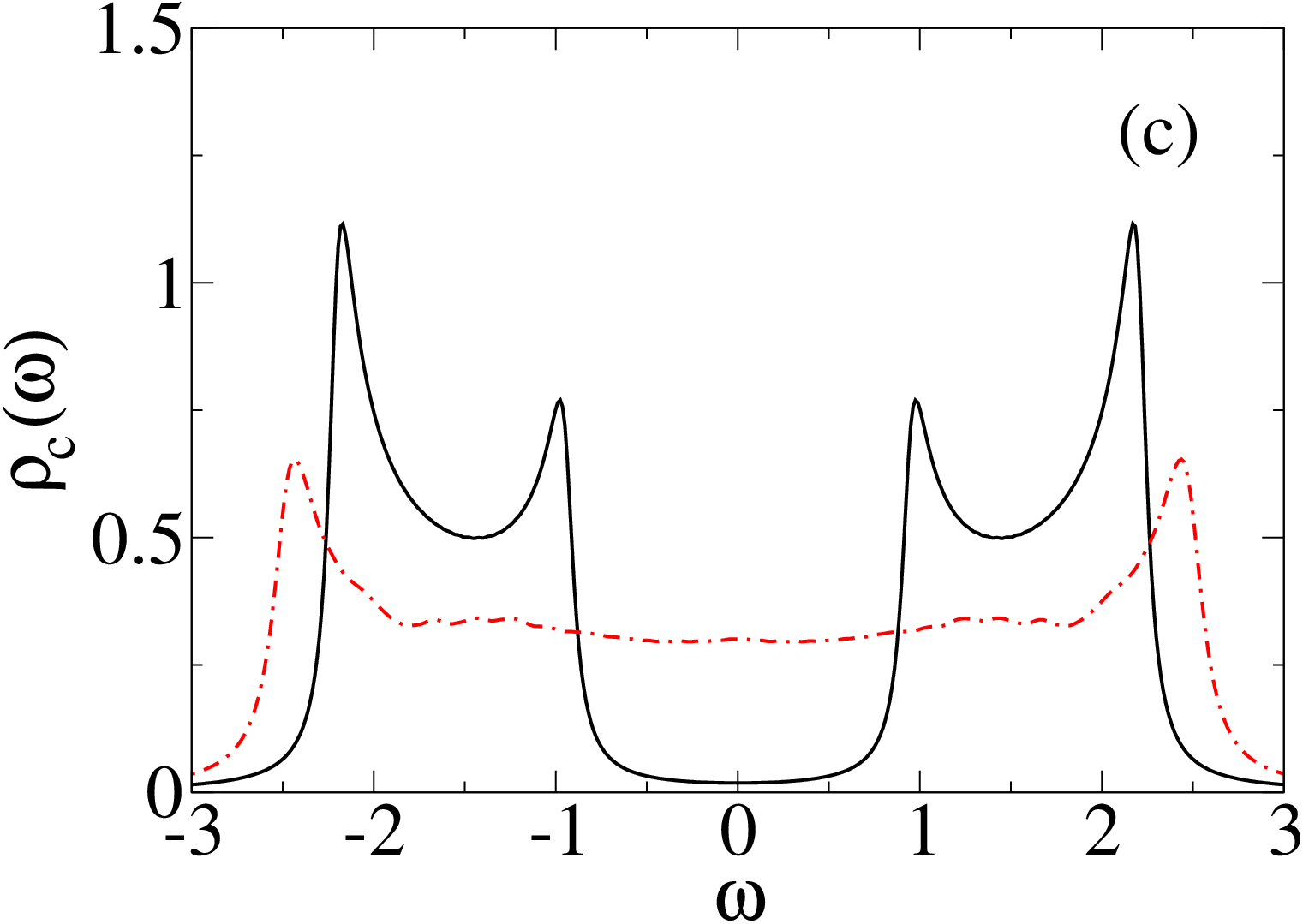}
\end{center}
\caption{(a) DOS $\rho_{c}(\omega)$ 
as function of $\omega$ 
for $U= 0.1$ (dotted line) and $U=1.0$ (solid line)
and $n_f=n_c=1/2$. Note the opening of a gap in the 
excitation spectrum at the Fermi level $\omega=0$. The dashed line is 
the exact result for $T=0$ as taken from  
Ref.~\cite{Lyzwa.93}. 
(b) DOS 
$\rho_{c}(\omega)$ for  a different filling $n_{f} = 1/3$ ($n_{c} = 2/3$)
for two values $U = 1.0$ (solid line) and $U = 2.0$ (dashed line). 
There are now two gaps found in the charge spectrum.
(c) Same quantity as in panel (a) with $n_f=n_c=1/2$
for a larger $U$ value $U=2.0$ (solid line). Besides the low temperature 
result the DOS is also shown for a higher temperature $T=0.2$ 
(dashed dotted line).
Note that the low-temperature gap at $\omega =0$ is smeared out for 
$T=0.2$.}
\label{fig:8}
\end{figure}

Next, we evaluate the dynamical properties using the PRM approach. 
First we consider the electronic density of states (DOS) 
of conduction electrons 
\begin{eqnarray}
\label{G79}
\rho_c(\omega) &=& \frac{1}{N} \sum_k (A_k^+(\omega)
+ A_k^-(\omega)) = \frac{1}{N} \sum_k
\frac{1}{2\pi} \int_{-\infty}^\infty  
\langle \; [\; c_k(t), c_k^\dagger\; ]_+ \;\rangle e^{i\omega t}dt.
\end{eqnarray}
where $A_k^+(\omega)$ and $A_k^-(\omega)$ were defined in 
\eqref{G56}.
The results from the PRM approach for the neutral case $n_f=n_c=1/2$ 
are shown in Figs.~\ref{fig:8}a and \ref{fig:8}c for three
different values of $U$.  For small $U$ the  
DOS resembles that of free conduction electrons in one dimension 
with a divergence at the lower and upper band edge (dotted line of
Fig.~\ref{fig:8}a). For larger $U$ the opening of a charge gap of
order $U$
around the Fermi level at $\omega=0$ can be observed 
(Fig.~\ref{fig:8}a and Fig.~\ref{fig:8}c). For still
larger $U$ only charge excitations away from the Fermi level are left. 
This behavior can easily be understood from the former expressions 
\eqref{G58}
for the two spectral functions. Note that for rather 
small values of $U$ the first terms in both equations 
are the dominant ones since the prefactors 
$|\tilde{\beta}_{\bf k k'}|^2$ of the second terms 
are of order $U^2$. From the gap in the renormalized 
quasiparticle energy $\varepsilon_k$ of Fig.~\ref{fig:5}a
the occurrence of a gap in $\rho_c(\omega)$  
immediately follows. Moreover, since the dispersion of 
$\varepsilon_k$ around the Fermi 
momentum flattens with increasing $U$ also the peaks in the DOS 
at the edges of the gap can be explained. The result for the larger 
$U$ value in Fig.~\ref{fig:8}c show a similar behavior as for the lower 
$U$'s. Note however, for these cases our weak-coupling approach might be at
the limit of its validity.
 
In Fig.~\ref{fig:8}a also the exact result for the 
DOS from Ref.~\cite{Lyzwa.93} at temperature $T =0$
is shown (dashed line). There, the divergencies 
at the band edges appear since an infinitely large system size was
considered in that approach. In contrast the renormalization equations
in the PRM approach were solved on a finite lattice by which the 
divergencies become smooth. On the other hand, in order to satisfy 
the sum rule the spectral function within the bands is enlarged
in the PRM result. In a recent Monte-Carlo study \cite{maska.05} 
the two-dimensional Falicov-Kimball model for the same concentrations 
$n_f=n_c=1/2$ was discussed. It shows  
quite similar features for the DOS as our PRM result for dimension $d=1$. 
In addition in Ref.~\cite{maska.05} also temperature-dependent
effects have been discussed.

In Fig.~\ref{fig:8}b the DOS $\rho_c(\omega)$ is shown for 
$n_f=1/3$ ($n_c=2/3$) for our lowest temperature 
$T= 0.1$. As expected from the previous  
results for $\tilde{\varepsilon}_k$ in Fig.~\ref{fig:5}b, 
there are now two gaps found in the charge excitation spectrum, 
one of which is again located at the Fermi level. Note that 
 in Fig.~\ref{fig:8}b the  density of states is symmetric 
with respect to the band center. This symmetry 
can easily be understood as follows: First,  
at low temperatures the correlation function 
$\big< \delta n_i \delta n_{i+\kappa}\big>$ 
is invariant with respect to translations by
three lattice sites (compare the lower panel of Fig.~\ref{fig:7}). 
Therefore, $C_\rho(k- k')
\sim N \; \delta_{k -k', \pm 2\pi/3}$ 
at low enough temperatures. By exploiting this relation 
and Eq.~\eqref{G36} one finds that all $k$ points which are 
symmetric to $\pm \pi/2$ lead to the same renormalization  
of the quasiparticle energy $\varepsilon_{k, \lambda}$ 
but with an opposite sign. Thus, the renormalized quasiparticle energy 
$\tilde{\varepsilon}_k$ becomes antisymmetric, 
since also the original $\varepsilon_k$ 
was antisymmetric relative to $\pm \pi/2$ and the
DOS becomes symmetric relative to the band center 
in the present approach. This result is in contrast to the expectation 
that for occupation $n_c$ away from half-filling 
the DOS should be asymmetric. In fact, 
one can assure oneself that 
higher order terms to  $X_{\lambda,\Delta \lambda}$ would give 
rise to a small additional asymmetric contribution to the DOS
in our approach. 
Note that such a small asymmetry is found 
in an exact result \cite{Lyzwa.93} for the DOS in one dimension
at $T=0$ for $n_c=2/3$.  

Finally, in  Fig.~\ref{fig:8}c also the influence of temperature on 
$\rho_c(\omega)$ is shown for  
$n_f=n_c=1/2$. Note that at $T=0.1$ the gap in the DOS is clearly seen.
However, by increasing the 
temperature by a factor of 2 the gap becomes completely smeared out. 
  \\

\begin{figure}[h]
\begin{center}
\includegraphics[width=5.2cm,height=4.2cm]{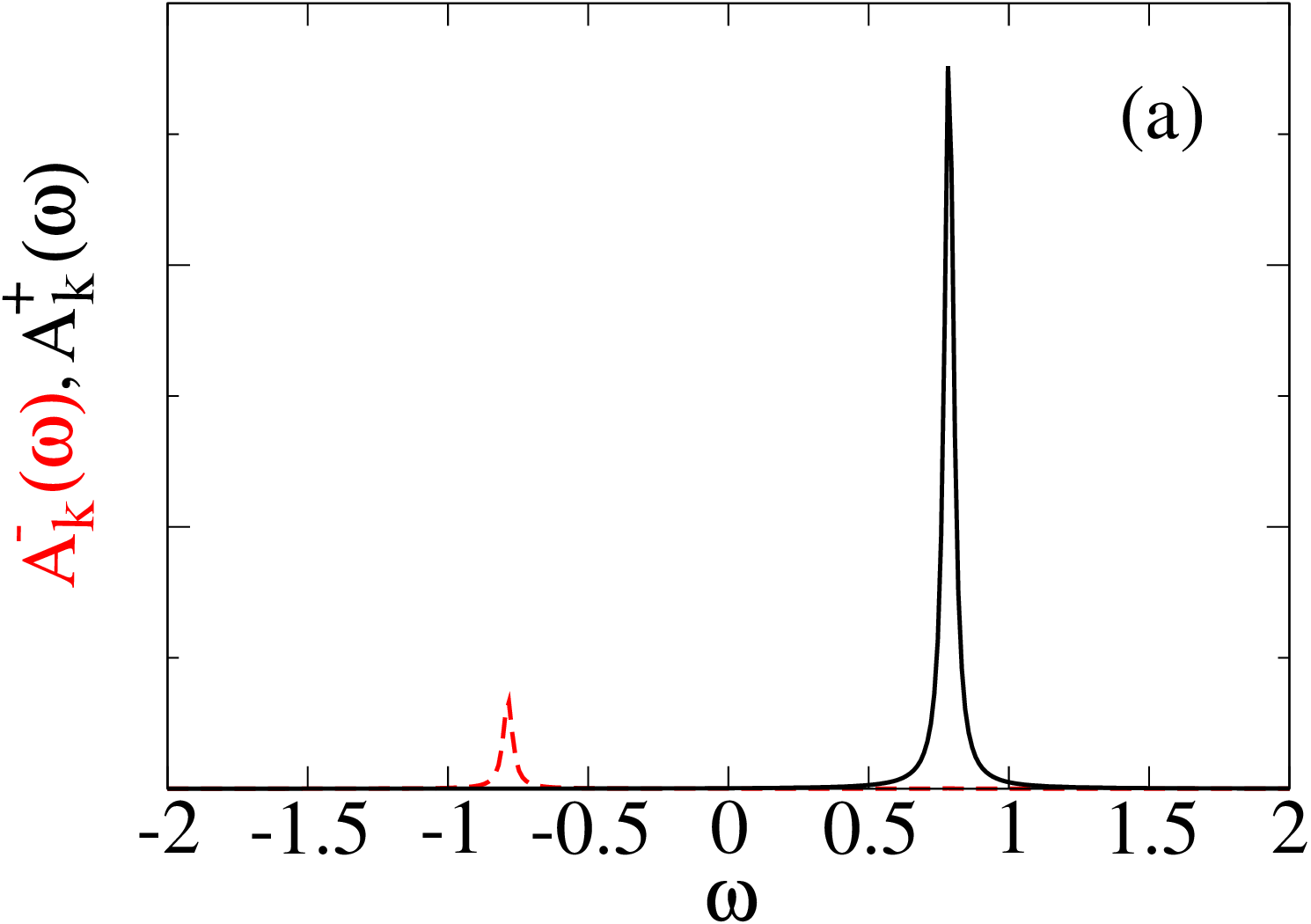} \hspace{0.1cm}
\includegraphics[width=5.2cm,height=4.2cm]{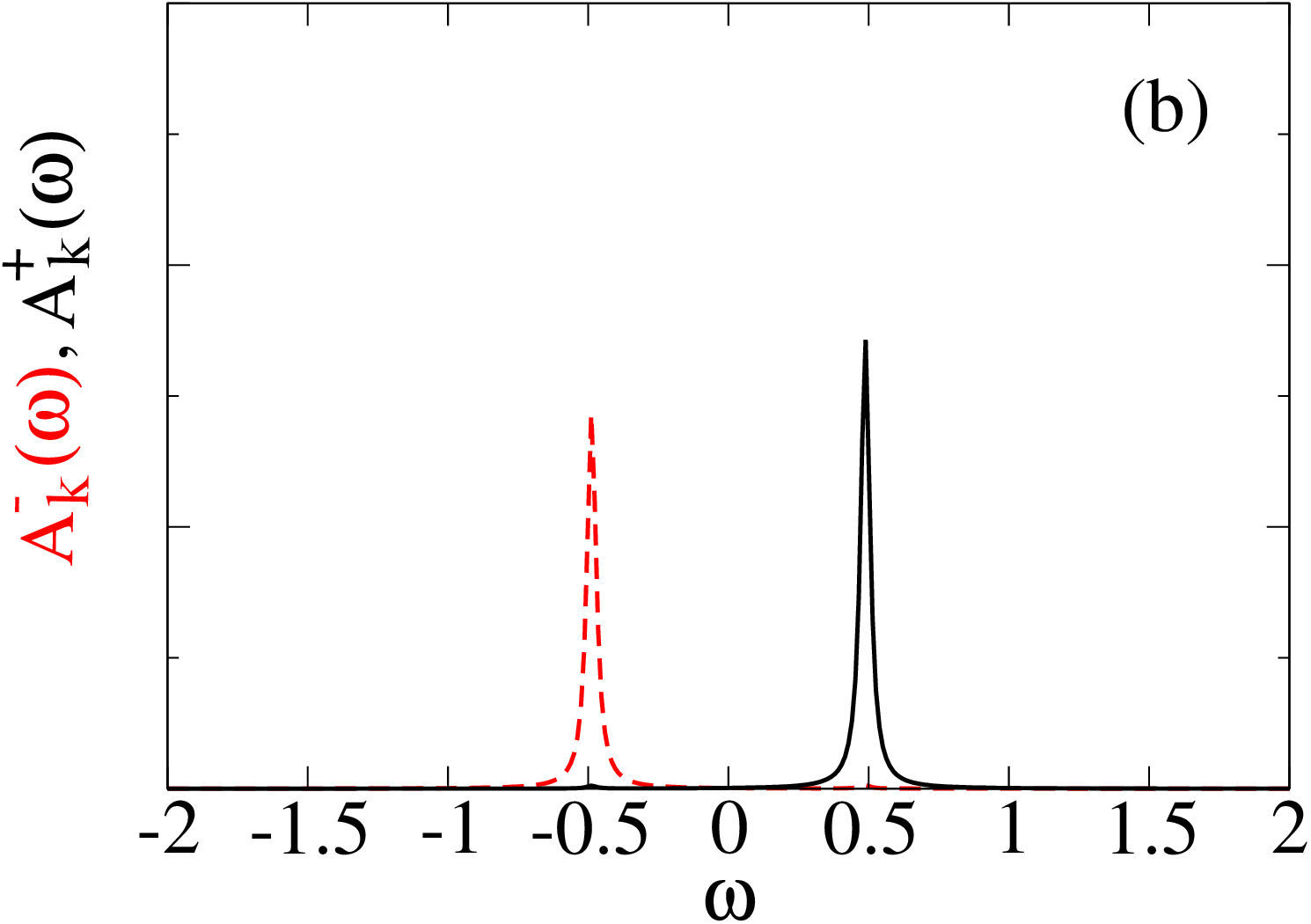} \hspace{0.1cm}
\includegraphics[width=5.2cm,height=4.2cm]{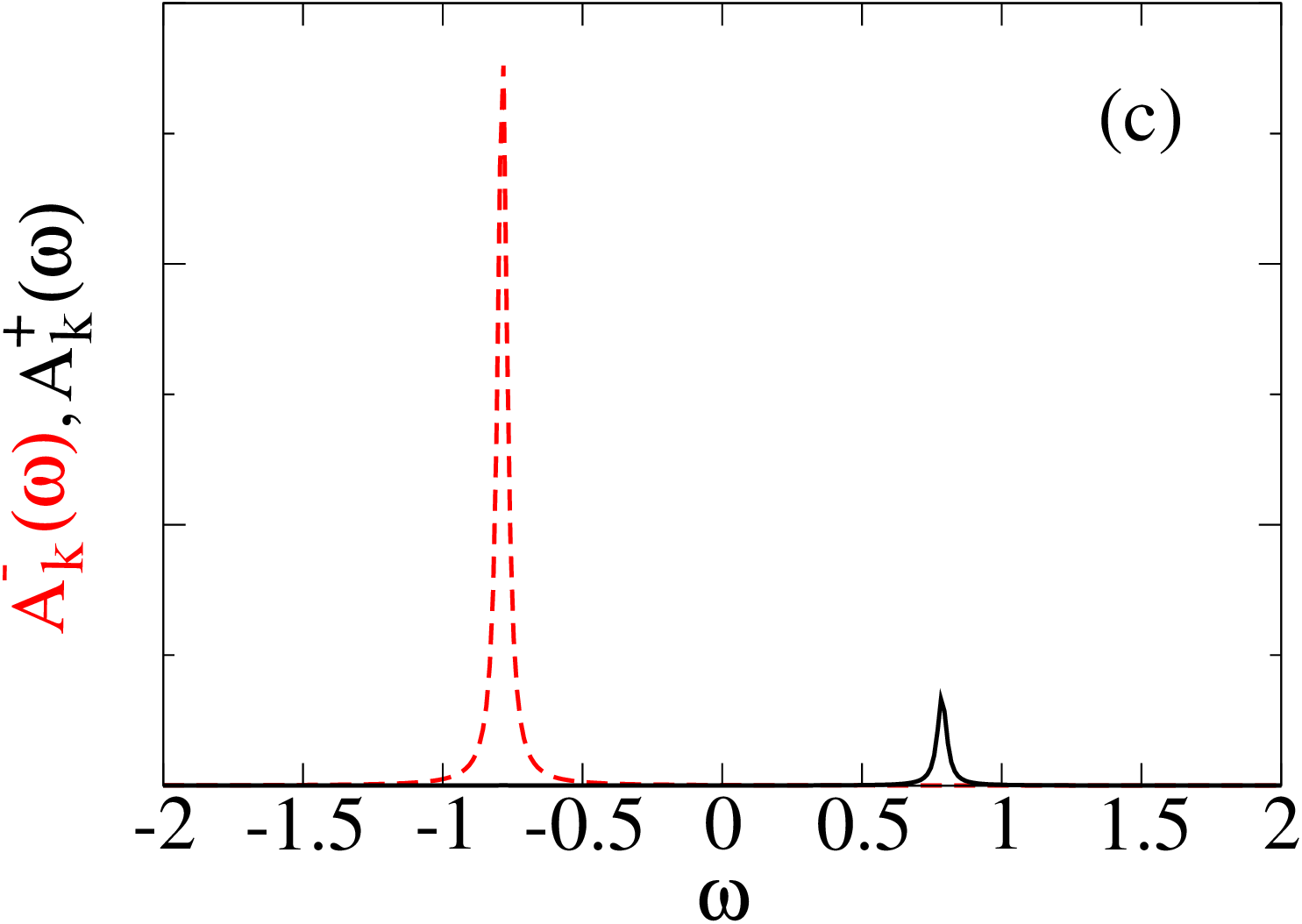}
\end{center}
\caption{One-particle spectral functions $A_k^+(\omega)$ (solid line) and 
$A_k^-(\omega)$ (dashed line) at low temperature ($T=0.1$) as function of 
$\omega$ for different $k$ vectors, 
(a) $k>k_F$, (b) $k  \ge k_F$, and (c) $k<k_F$. ($U=1.0$, $n_f=n_c= 
1/2$.)}
\label{fig:14}
\end{figure}

In Figs.~\ref{fig:14}a-c the one-particle spectral functions 
$A_k^+(\omega)$ and  $A_k^-(\omega)$ for different $k$ values
are shown for the lowest possible temperature 
of $T=0.1$ and $n_f=n_c=1/2$. For $k > k_F$ (Fig.~\ref{fig:14}a) a
coherent excitation is found in $A_k^+(\omega)$ 
and a small incoherent excitation in $A_k^-(\omega)$.
To understand
the latter feature, note that only a single $k'$ value, 
$k'=k - \pi$, contributes to the incoherent part  in \eqref{G58} due to
$C_\rho(k-k') \sim N \delta_{k-k',\pi}$.  The same remains true  
when $k$ comes closer to $k_F$ (Fig.~\ref{fig:14}b). In this case, however, 
the incoherent contribution from $A_k^-(\omega)$ has gained 
weight. The reason is the renormalization factor $|\tilde{\beta}_{kk'}|^2$ 
in equation \eqref{G58}. It becomes stronger when $k$ and also 
$k'$ approaches the Fermi momentum as follows from \eqref{G47} 
and \eqref{G30}. 
Finally, for $k<k_F$   (Fig.~\ref{fig:14}c)
a coherent contribution is found in  $A_k^-(\omega)$ and an 
 incoherent excitation in $A_k^+(\omega)$.

\begin{figure}[h]
\begin{center}
\includegraphics[width=5.4cm,height=4.2cm]{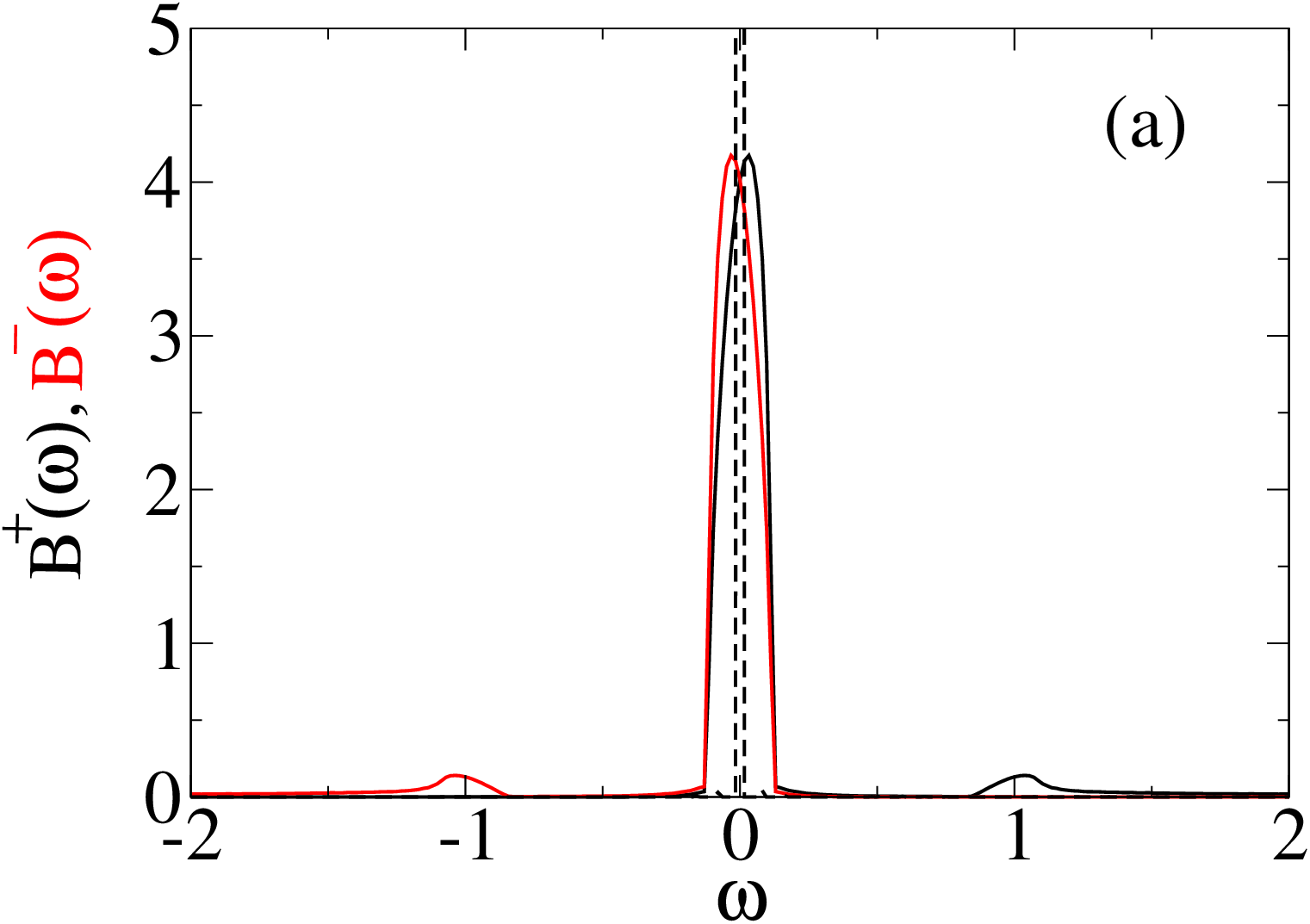}
\includegraphics[width=5.4cm,height=4.2cm]{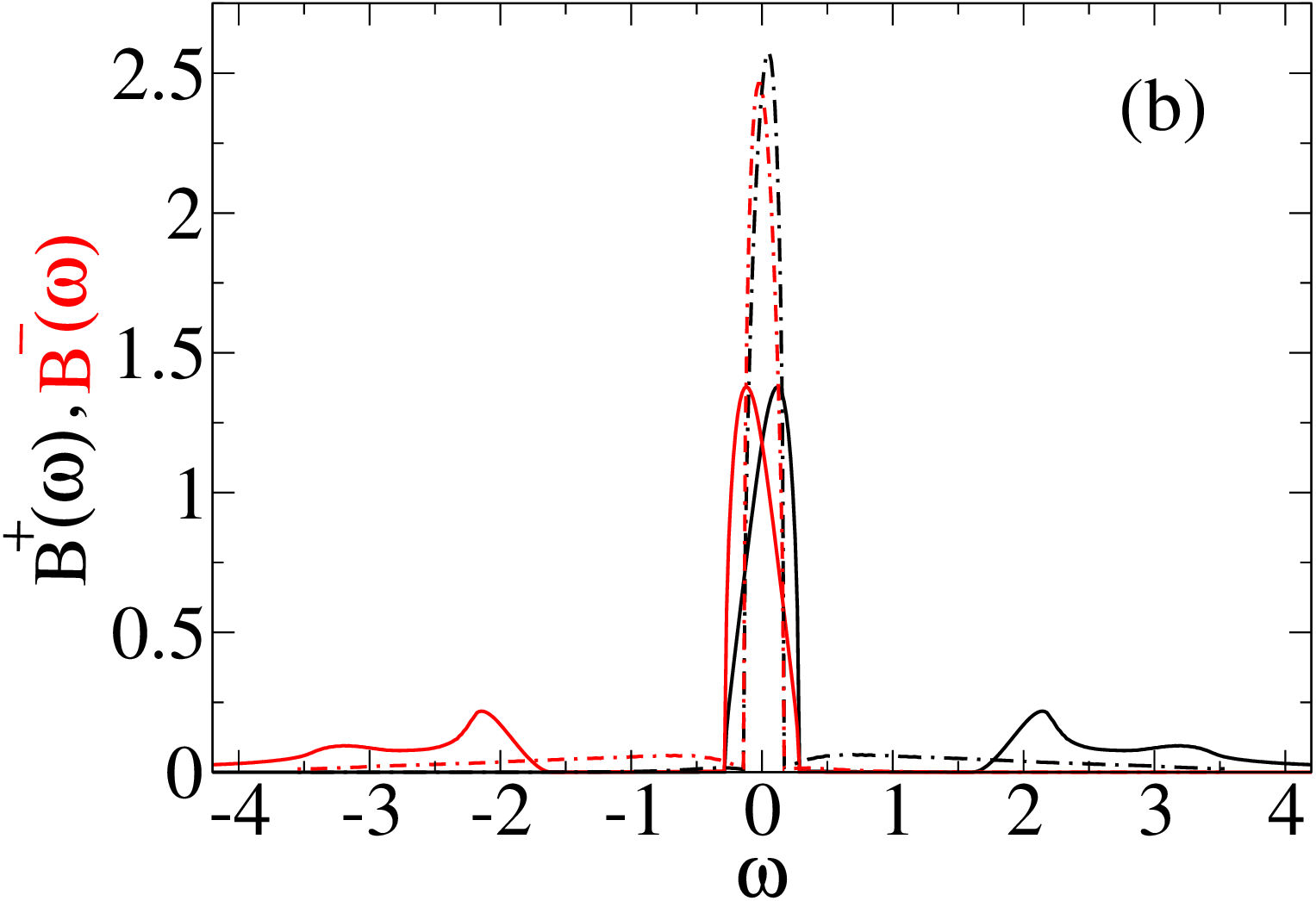}
\includegraphics[width=5.4cm,height=4.2cm]{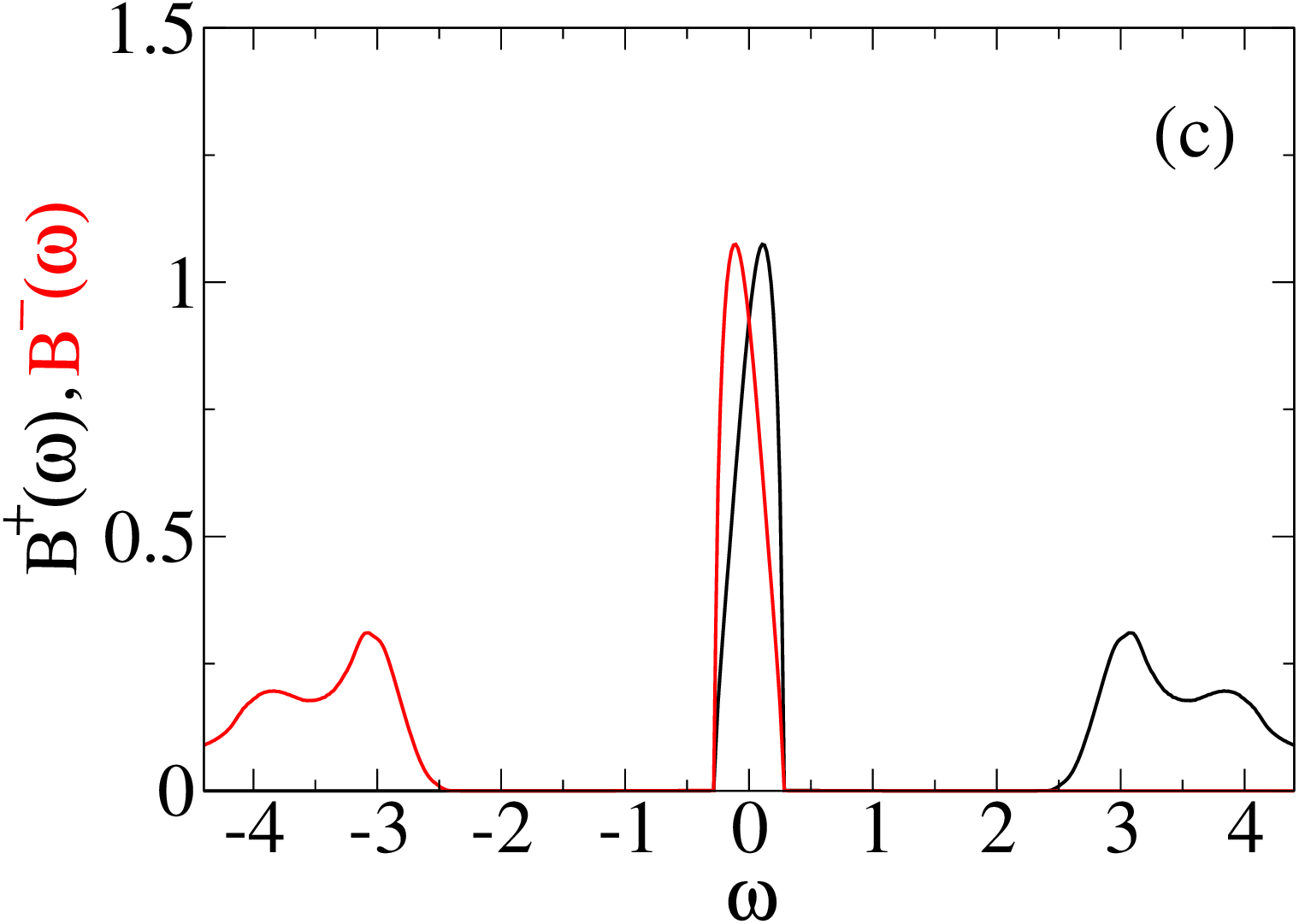}
\end{center}
\caption{(a) One-particle spectral functions
$B^{+}(\omega)$ (black) and $B^{-}(\omega)$ (red) of
 $f$-electrons as function of $\omega$
for $U=0.1$ (dashed line) and $U=1.0$ (solid line),   
  and $n_f=n_c=1/2$, $T=0.1$. For both $U$ values the coherent 
 excitation dominates the spectrum. (b),(c) Same quantity as in 
Fig.~\ref{fig:11}a
for two larger values of $U$, $U=2.0$ (b) and $U=3.0$ (c)
and $T=0.1$ (solid lines). Note that now the incoherent 
excitations have gained considerable weight 
as compared to the case of panel (a).
For $U=2.0$ (b) the
spectral function is also shown for a higher temperature $T=0.2$ 
(dashed dotted line).
\vspace*{1cm}}

\label{fig:11}
\end{figure}

In Figs.~\ref{fig:11}a-c the one-particle 
spectral functions $B^{+}(\omega)$ (black) and $B^{-}(\omega)$ (red)
of the localized electrons averaged over all 
lattice sites from section IV are shown for different values of
$U$ and $n_f=n_c=1/2$.  For the two lower $U$-values ($U=0.1, 1.0$) 
at a rather low temperature $T=0.1$ the coherent parts (of weight
$\tilde{\nu}^2$) dominate in Fig.~\ref{fig:11}a
the spectrum (compare \eqref{G61}). 
In contrast, for the two larger $U$-values ($U=2.0, 3.0$) 
the incoherent excitations 
have gained considerable weight (Fig.~\ref{fig:11}b and Fig.~\ref{fig:11}c).   
Note that at a higher temperature $T=0.2$ the
incoherent part is smeared out over a wide frequency range 
for $U=2.0$ (dashed dotted curve in
Fig.~\ref{fig:11}b). The same feature would also be expected for $U=3.0$. 
However, no stable solution of the renormalization equations 
was found for this temperature. The discussed behavior can again be 
understood from the former PRM expressions \eqref{G65}, \eqref{G66} 
for the spectral functions $B^{+}(\omega)$ and $B^{-}(\omega)$.
The first term in both equations describes the coherent excitation
of weight $\tilde{\nu}^2$ at $\omega \approx \tilde{\varepsilon}_f =0$
whereas the second term follows from the coupling of $f_i$ to
electronic particle-hole excitations (compare \eqref{G61}). Their weight 
$\tilde{\mu}_{k,k'}^2$ is at least of order $U^2$. The gaps 
in  Figs.~\ref{fig:11}a-c of order $U$ immediately follow 
from the gap in the quasiparticle energy $\tilde{\varepsilon}_k$.

\begin{figure}
\begin{center}
\includegraphics[width=8cm,height=6cm]{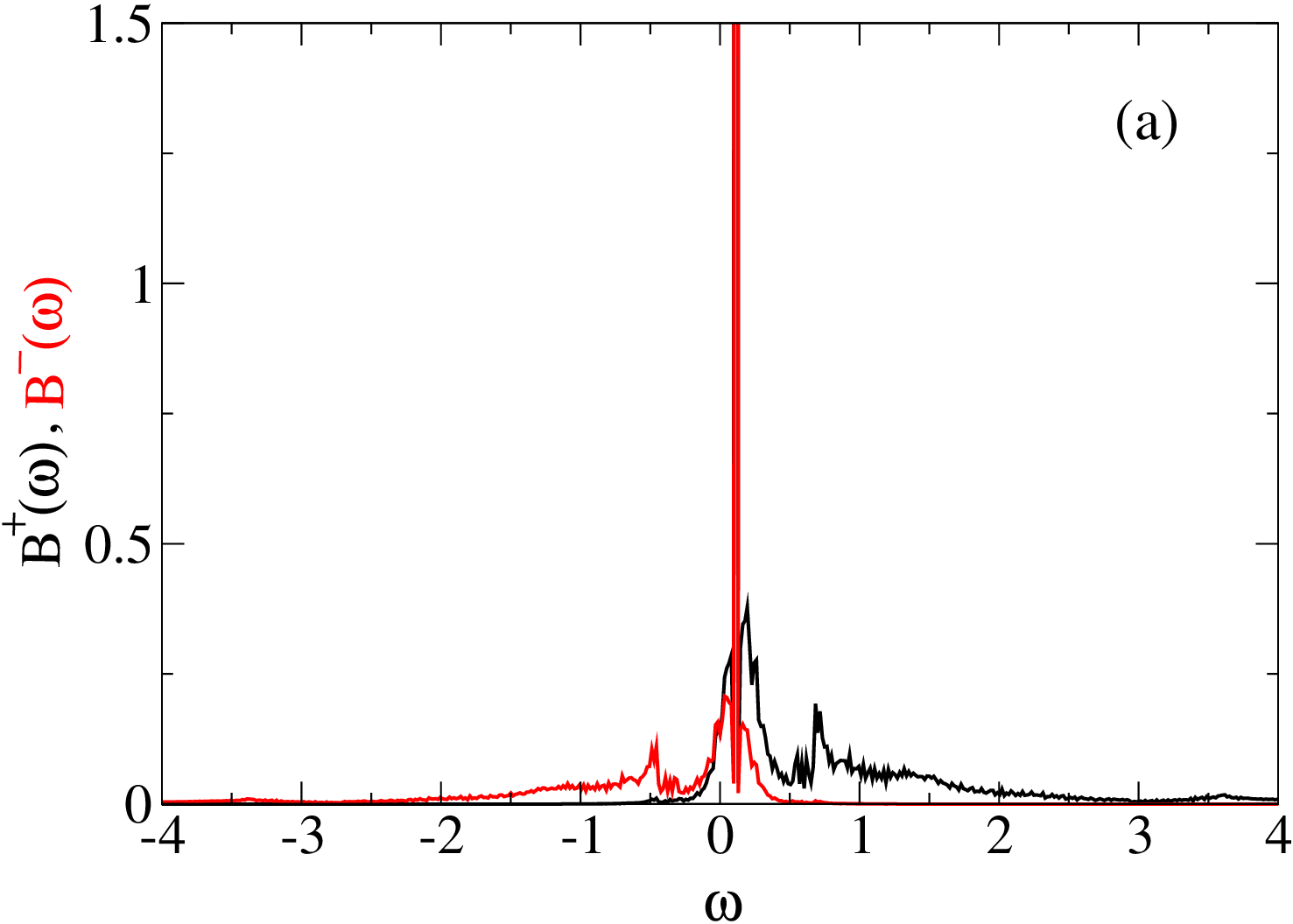}
\includegraphics[width=8cm,height=6cm]{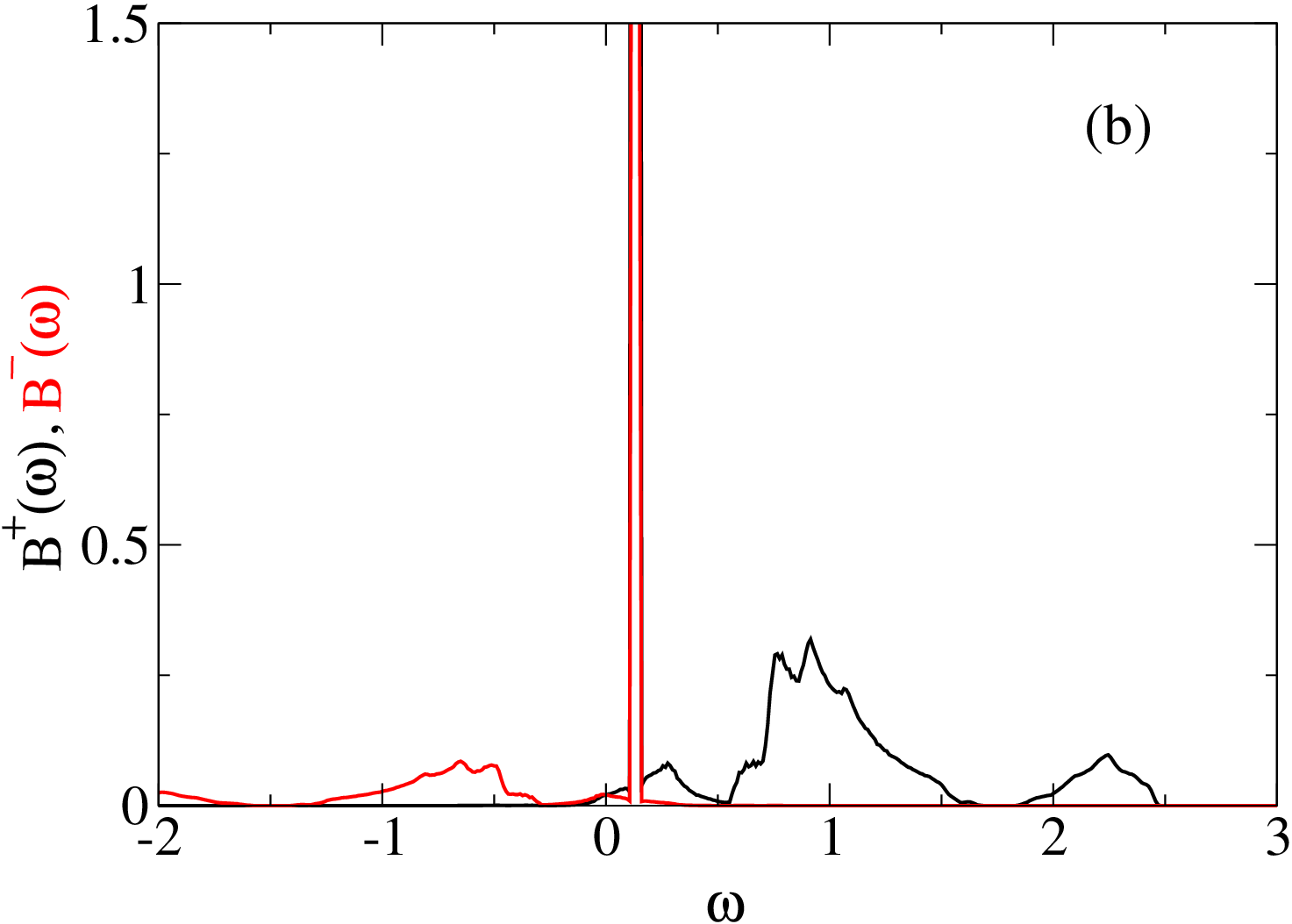}
\end{center}
\caption{Same quantity as in Fig.~\ref{fig:11} 
for $U=1.0 $ (a) and $U=2.0 $ (b)   
  and $n_f=1/3$.}
\label{fig:13}
\end{figure}

The same quantities 
 $B^{+}(\omega)$ (black) and $B^{-}(\omega)$ (red) are also shown in 
Fig.~\ref{fig:13}  for $f$-filling $n_f=1/3$ $(n_c=2/3)$ and  $U= 1.0,
2.0$. The gaps between the respective coherent and incoherent
parts again correspond to the gap  
of the quasiparticle excitation in Fig.~\ref{fig:5}b 
at the Fermi level. Note that the spectrum is no
longer symmetric around $\omega = 0$. The different 
weights of  $B^{+}(\omega)$ and $B^{-}(\omega)$ in Fig.~\ref{fig:13} 
follow from the Fermi functions in \eqref{G65} and \eqref{G66}.    

We have also checked the $f$-sum rule 
\begin{eqnarray}
\label{G80}
\int_\infty^\infty d \omega \;
(B^+(\omega) + B^-(\omega)) &=& \big< [f_i^\dagger, f_i ]_+ \big> =1
\end{eqnarray} 
for the spectral functions in Figs.~\ref{fig:11} and 
\ref{fig:13}. We found that for the two  
small $U$ values in Fig.~\ref{fig:11}a
the sum rule was perfectly fulfilled whereas for the spectral 
functions with $U=2.0$ and 
$3.0$ (Fig.~\ref{fig:11}b and Fig.~\ref{fig:11}c) the agreement was less satisfactory. 
For instance, for $U=3.0$ a value of $0.9$ instead of $1$ was found.   
The origin of these deviations 
can be understood as follows. First, from Eqs.~\eqref{G65} and
\eqref{G66} one finds that \eqref{G80} is fulfilled if 
\begin{eqnarray}
\label{G81}
|\nu_\lambda|^2 + \frac{1}{N^2} \sum_{k k'}
|\mu_{k k',\lambda}|^2 f({\varepsilon}_{k, \lambda})
(1- f(\varepsilon_{k',\lambda})) =1 
\end{eqnarray} 
is satisfied for all values of $\lambda$. For the original model  
relation \eqref{G81} is trivially valid. By 
restricting oneself to the lowest 
order processes in the renormalization equations 
for the $\lambda$-dependent coefficients in \eqref{G61}
one can show  that the sum rule is fulfilled for $(\lambda -
\Delta \lambda)$ if it was fulfilled for $\lambda$. However, 
there are deviations due to higher order renormalization 
contributions for larger $U$ and small values of $\lambda$ 
as follows from expression \eqref{G30} for $A_{k,k'}^{\lambda, 
\Delta \lambda}$. Also important for the deviations 
is the chosen simple operator structure of the 
generator $X_{\lambda, \Delta \lambda}$
from \eqref{G16}. Note however that higher 
order processes in $U$ are suppressed for larger system sizes $N$. 
Thus, in order to control the influence of higher order terms 
a finite size scaling in $N$ should be done. Instead, in the present
approach a simpler truncation procedure has been used. The renormalization 
was stopped at the smallest $\lambda$ value at which the influence of 
higher order terms was still negligible.  
This leads to the observed difference in the sum rule as compared 
to the exact result. 
 \vspace{0.5cm}

\section{Conclusions
\label{Conclusions}}
In summary, in this paper we have applied a recently developed
projector-based renormalization method (PRM) to the spinless
Falicov-Kimball model in one dimension and at half-filling
$n_f +n_c=1$. 
As a starting point in the formalism, we have taken 
the one-particle part of the initial Hamiltonian \eqref{G1} as
unperturbed Hamiltonian ${\cal H}_0$ and  
the local Coulomb interaction $U$  
between conduction and localized electrons as interaction ${\cal H}_1$.
The elimination of the latter part leads to the
renormalization of the Hamiltonian. Thereby a new density-density
interaction between the localized electrons is generated. Finally, a
number of static and dynamic quantities has been calculated.
Note that the present formuation of the PRM is a weak-coupling
approach which however should lead to reliable results for 
$U$-values up to the order of $t$. It is shown that 
the investigation of the small temperature behavior 
enables us also to extract valuable information 
about the zero temperature properties of static and dynamic
quantities.  

\section*{Acknowledgments}
We would like to acknowledge fruitful discussions with J.~Freericks, 
Ch.~Geibel and 
A.~H\"ubsch. This work was supported by the 
DFG through the research program  SFB 463, and by DFG grant 
No: Dr 274/10-1. One of us (VZ) would like to acknowledge the support by the
National Science Foundation under grant number DMR-0210717.

\newpage

\begin{appendix}
\section{Operator transformation of ${\cal H}_\lambda$}
\label{trafos}
In this appendix we evaluate the renormalized Falicov-Kimball Hamiltonian  
obtained by the unitary transformation 
${\cal H}_{(\lambda -\Delta\lambda)}
= 
\exp\{X_{\lambda, \Delta \lambda}\} \ {\cal H}_{\lambda} \ 
\exp\{-X_{\lambda, \Delta \lambda}\},
$
where $X_{\lambda, \Delta \lambda}$ is given by \eqref{G22}
The transformed Hamiltonian ${\cal H}_{\lambda -\Delta \lambda}$
has no excitations with transfer energies 
$|\varepsilon_{k, \lambda - \Delta \lambda} -
\varepsilon_{k', \lambda - \Delta \lambda}| > 
\lambda - \Delta \lambda $. Let us first consider 
the transformation of ${\cal H}_{0,\lambda}$
\begin{eqnarray}
\label{A1}
e^{X_{\lambda, \Delta \lambda}} {\cal H}_{0,\lambda}
e^{-X_{\lambda, \Delta \lambda}} - {\cal H}_{0,\lambda}
&=& \sum_k \varepsilon_{k, \lambda}
\sum_{n=1}^\infty \frac{1}{n!}\; 
{\bf X}_{\lambda, \Delta \lambda}^n c_k^\dagger c_k
\end{eqnarray}
where we have introduced a new superoperator 
${\bf X}_{\lambda, \Delta \lambda}$ which is defined by the commutator of the
generator $X_{\lambda, \Delta \lambda}$ with operators ${\cal A}$ on
which ${\bf X}_{\lambda, \Delta \lambda}$ is applied, 
${\bf X}_{\lambda, \Delta \lambda} {\cal A} = 
[X_{\lambda, \Delta \lambda}, {\cal A}]$. Note that the operators 
$c_k^\dagger c_k$ are the only part of ${\cal H}_{0,\lambda}$ 
which do not commute with $X_{\lambda, \Delta \lambda}$. We
successively evaluate the commutators on the right-hand side of
\eqref{A1}. By use of  
\begin{eqnarray}
\label{A2}
{\bf X}_{\lambda, \Delta \lambda} c_k^\dagger c_k
&=& 
- \frac{1}{N} \sum_{\tilde{k},i} A_{k \tilde k}
\; \Theta_{k \tilde k}^{\lambda, \Delta \lambda} \;
\Big( e^{i(\tilde k- k){R}_i}\; \delta n_i \;
c_{\tilde k}^\dagger c_k +
e^{i(k- \tilde k){R}_i}\; \delta n_i \;
c_{k}^\dagger c_{\tilde k} 
\Big)
\end{eqnarray}
we first obtain 
\begin{eqnarray}
\label{A3}
{\bf X}_{\lambda, \Delta \lambda} {\cal H}_{0,\lambda}
&=& 
\frac{1}{N} \sum_{kk'i} A_{k k'}
\; \Theta_{k k'}^{\lambda. \Delta \lambda} \;
(\varepsilon_{k',\lambda} - \varepsilon_{k, \lambda}) \;
e^{i(k- k'){R}_i}\; \delta n_i \;
c_{k}^\dagger c_{k'} 
\end{eqnarray}
Next  by applying ${\bf X}_{\lambda, \Delta \lambda}$ on 
${\bf X}_{\lambda,\Delta \lambda}{\cal H}_{0,\lambda}$ 
where products of two fermions and of two
local density operators $\delta n_j \; \delta n_i$ occur. In order to keep 
only operators which are also present in 
${\cal H}_{(\lambda - \Delta \lambda)}$ a factorization approximation 
has to be performed. With 
\begin{eqnarray}
\label{A4}
 {\bf X}_{\lambda, \Delta \lambda} \;
 \delta n_i \; c_{k}^\dagger c_{k'} &=& 
\frac{1}{N} \sum_{\tilde{k},j} 
A_{\tilde{k} k} \Theta_{\tilde{k} k}^{\lambda, \Delta \lambda} \;
e^{i(\tilde{k}- k){R}_j}\;
\Big( \big< \delta n_i \delta n_j \big> c_{\tilde k}^\dagger c_{k'}
\\
&& \hspace*{2cm} 
+ \delta_{k', \tilde k} \; 
\delta n_i \delta n_j  \big< c_{k'}^\dagger c_{k'} \big>
- \delta_{k', \tilde k}
\big< \delta n_i \delta n_j \big> \big<c_{\tilde k}^\dagger 
c_{k'} \big> 
\Big) \nonumber \\
&-& \frac{1}{N} \sum_{\tilde k,j} 
A_{\tilde{k} k'} \Theta_{\tilde{k} k'}^{\lambda, \Delta \lambda} \;
e^{-i(\tilde{k}- k'){R}_j}\;
\Big( \big< \delta n_i \delta n_j \big> c_{k}^\dagger c_{\tilde{k}}
\nonumber \\
&& \hspace*{2cm} 
+ \delta_{k, \tilde{k}} \; 
\delta n_i \delta n_j  \big< c_{k}^\dagger c_k \big>
- \delta_{k, \tilde{k}}
\big< \delta n_i \delta n_j \big> \big<c_{k}^\dagger 
c_{\tilde{k}} \big> 
\Big) \nonumber 
\end{eqnarray}
one finds 
\begin{eqnarray}
\label{A5}
{\bf X}_{\lambda, \Delta \lambda}^2 {\cal H}_{0,\lambda}
&=&
\frac{2}{N^2}
\sum_{k k'ij} A_{k k'}^2 \Theta_{kk'}^{\lambda, \Delta \lambda} 
(\varepsilon_{k',\lambda} - \varepsilon_{k,\lambda})
\cos[i(k'-k)({R}_i -{R}_j)] \\
&& \times \Big( \big< \delta n_i \delta n_j \big> 
c_{\tilde{k}}^\dagger c_{k'} \nonumber
+ \delta_{k', \tilde{k}} \; 
\delta n_i \delta n_j  \big< c_{k'}^\dagger c_{k'} \big>
- \delta_{k', \tilde{k}}
\big< \delta n_i \delta n_j \big> \big<c_{\tilde{k}}^\dagger 
c_{k'} \big> 
\Big) \nonumber 
\end{eqnarray}
or by using translational invariance
\begin{eqnarray}
\label{A6}
{\bf X}_{\lambda, \Delta \lambda}^2 {\cal H}_{0,\lambda}
&=&
\frac{2}{N}
\sum_{k k'}  A_{k k'}^2 \Theta_{kk'}^{\lambda, \Delta \lambda} 
(\varepsilon_{k',\lambda} - \varepsilon_{k,\lambda}) \\
&\times& \Big( C_\rho(k' -k)
c_{k}^\dagger c_k \nonumber
+ \big< c_{k}^\dagger c_k \big>
\frac{1}{N} \sum_{ij} \delta n_i \delta n_j\; e^{i(k- k')
({R}_i - {R}_j)}  
-  \big<c_k^\dagger c_k \big> C_\rho(k'-k)  
\Big) \nonumber 
\end{eqnarray}
Note that in deriving \eqref{A5} and \eqref{A6} it was assumed that the number
of $k$ points which are integrated out in the step  from 
$\lambda$ to $(\lambda - \Delta \Lambda)$ is small compared to the total
number of $k$ points. Therefore, the product of two 
product $\delta$ functions 
$\Theta_k k'$ and $\Theta_{\tilde{k} \tilde{k}'}$ is only  
nonzero when the wavevectors are pairwise equal to each other
(shell condition).  
In the next step, ${\bf X}_{\lambda, \Delta \lambda}$ is again 
applied on $c_k^\dagger c_k$. Thus by use of \eqref{A2}
one finds  
\begin{eqnarray}
\label{A7}
{\bf X}_{\lambda, \Delta \lambda}^3 {\cal H}_{0,\lambda}
&=& 
\frac{1}{N} \sum_{kk'i} 
\big[ -\frac{2^2 C_\rho(k -k')A_{k k'}^2}{N} \big]
A_{k k'}
\; \Theta_{k k'}^{\lambda, \Delta \lambda} \;
(\varepsilon_{k',\lambda} - \varepsilon_{k, \lambda}) \;
e^{i(k- k'){R}_i}\; \delta n_i \;
c_{k}^\dagger c_{k'} \nonumber \\
&& 
\end{eqnarray}
which has again the structure of ${\bf X}_{\lambda, \Delta \lambda}
{\cal H}_{0,\lambda}$ except of the additional prefactor in the bracket
$ [ ... ]$. thus one may sum up all commutator terms from \eqref{A1}
in order to obtain a compact expression for \eqref{A1}. 
In the same way one may proceed 
with the transformation of ${\cal H}_{1,\lambda}$.

\section{Green function ${\cal G}_0(\omega)$
of the lattice gas model}
To evaluate the Green function ${\cal G}_0(\omega)$ 
\begin{eqnarray}
\label{C1}
 {\cal G}_0(\omega) &=& i \int_0^\infty dt \; e^{i(\omega + i\eta)t}\;
\big< [f_i(t), f_i^\dagger]_+ \big>_{\tilde{\cal H}_f}
\end{eqnarray}
we best use the Mori-Zwanzig projection formalism. First, let us define
an anticommutator scalar product for  
operators ${\cal A}$, ${\cal B}$ by
$ ({\cal A} | {\cal B} ) = \big< [{\cal A}^\dagger , {\cal B} ]_+
\big>_{\tilde{\cal H}_f}$. 
In addition, we introduce the Liouville operator $\tilde{\bf L}_f$  
belonging to the lattice gas Hamiltonian $\tilde{\cal H}_f$. It acts 
on any operator ${\cal A}$ of the unitary space as  
$\tilde{\bf L}_f{\cal A} = [{\cal H}_f, {\cal A}]$. 
Thus \eqref{C1} can be rewritten as
\begin{eqnarray}
\label{C3}
{\cal G}_0(\omega) &=& i \int_0^\infty dt \;
e^{i(\omega +i \eta)t}\; \big(f_i^\dagger| e^{-i \tilde{\bf L}_f\;t } \;
f_i^\dagger \big) = 
\big( f_i^\dagger | \frac{1}{\tilde{\bf L}_f - \omega - i\eta}\;
f_i^\dagger \big)
\end{eqnarray}
The well-known operator identity of the projection formalism \cite{fick.90}
can be used to transform ${\cal G}_0(\omega)$ into 
\begin{eqnarray}
\label{C4}
{\cal G}_0(\omega) &=& \frac{1}{\Omega -\omega - \Sigma(\omega)}
\end{eqnarray}
where the frequency term $\Omega$ and the selfenergy $\Sigma(\omega)$
are given by
\begin{eqnarray}
\label{C5}
\Omega &=& \big( f_i^\dagger| \tilde{\bf L}_f f_i^\dagger \Big)
\hspace*{2cm} 
\Sigma(\omega) = \big( \tilde{\bf L}_f f_i^\dagger| {q}
\frac{1}{ {\bf Q} \tilde{\bf L}_f {\bf Q} - \omega - i \eta}\;
{\bf Q} \tilde{\bf L}_f f_i^\dagger \big)
\end{eqnarray}
Note that in \eqref{C4} and \eqref{C5} we have already used that 
$\big( f_i^\dagger| f_i^\dagger \Big) 
= \big< [ f_i, f_i^\dagger]_+ \big>_{\tilde{\cal H}_f} =1 $.
The quantity ${\bf Q}$ is like ${\bf L}_f$ a superoperator which acts on 
usual operators of the unitary space. ${\bf Q}$ is a projection operator 
and projects on the subspace of the Liouville space 
perpendicular to $|f_i^\dagger\big)$, i.e., it is defined by
${\bf Q} |{\cal A}\big) = |{\cal A} \big)
- | f_i^\dagger \big) \big( f_i^\dagger|{\cal A}\big)$. To evaluate
$\Omega$ and $\Sigma(\omega)$ we need
\begin{eqnarray}
\label{C6}
\tilde{\bf L}_f |f_i^\dagger \big) &=&
\tilde{\varepsilon}_f |f_i^\dagger \big) +
2 \sum_{j(\neq i)} \tilde{g}_{ij} |\delta n_j f_i^\dagger \big)
\hspace*{0.5cm} \mbox{and} \hspace*{0.5cm}
{\bf Q} \tilde{\bf L}_f |f_i^\dagger \big) = 
2 \sum_{j(\neq i)} \tilde{g}_{ij} {\bf Q}|\delta n_j f_i^\dagger \big)
\end{eqnarray}
where in the second relation the first term of 
$\tilde{\bf L}_f |f_i^\dagger \big)$ drops due 
to the presence of ${\bf Q}$. As is easily seen, we obtain 
$\Omega = \tilde{\varepsilon}_f$ and  
\begin{eqnarray}
\label{C7}
\Sigma(\omega) &=& 
4 \sum_{j(\neq i)} \sum_{l (\neq i)} \tilde{g}_{il}^* \; \tilde{g}_{ij} \;
\big( \delta n_l f_i^\dagger |{\bf Q} 
\frac{1}{ {\bf Q} \tilde{\bf L}_f {\bf Q} - \omega - i \eta}\;
{\bf Q} |\delta n_j f_i^\dagger \Big)
\end{eqnarray}
Note that 
the local density operators $\delta n_l$ and $\delta n_j$ 
in the bra- and ket vector of \eqref{C7} 
commute with $\tilde{\cal H}_f$. Therefore, we can
approximate $\Sigma(\omega)$ by an expression proportional to 
${\cal G}_0$ 
\begin{eqnarray}
\label{C8}
\Sigma(\omega) &=& \kappa^2 \; {\cal G}_0(\omega) 
\hspace*{2cm}
\kappa^2 =  
4 \sum_{j(\neq i)} \sum_{l (\neq i)} \tilde{g}_{il}^* \; \tilde{g}_{ij} \;
\big< \delta n_l \delta n_j \big>_{\tilde{\cal H}_f} 
\end{eqnarray}
which is identical to \eqref{G67}. Note that 
$\kappa^2$ in \eqref{C8} is independent of the lattice site $i$. 
$\Sigma(\omega)$ together with \eqref{C4} forms a quadratic 
equation for ${\cal G}_0(\omega)$ which can be solved. 
It is easily seen that the imaginary part
is given by  
\begin{eqnarray}
\label{C9}
\Im   {\cal G}_0(\omega) &=& \frac{1}{\kappa} 
\sqrt{1 - \frac{(\omega - \tilde{\varepsilon}_f)^2}{4 \kappa^2}}
\end{eqnarray}
for the frequency range $|\omega - \tilde{\varepsilon}_f| \leq 2 \kappa$ 
and ${\cal G}_0(\omega)=0$ elsewhere.
Thus, 
$\Im {\cal G}_0(\omega)$ is nonzero 
only in a small frequency region around $\omega =\tilde{\varepsilon}_f$ of
width $\pm 2\kappa$ where $\kappa$ is determined by static density-density 
correlations of the lattice gas model. 
The maximum value of $\Im {\cal G}_0(\omega)$ is located 
at $\omega= \tilde{\varepsilon}_f$ and is given by $1/\kappa$. 

From  $\Im{\cal G}_o(\omega)$ the functions $B_0^+(\omega)$ 
and $B_0^-(\omega)$ are determined by help of the  
fluctuation-dissipation theorem 
\begin{eqnarray}
                   \label{C10}
B_0^+(\omega) &=& (1 -f(\omega)) \frac{1}{\pi}
\Im {\cal G}_0(\omega) 
\hspace*{2cm} 
B_0^-(\omega) = f(\omega) \frac{1}{\pi} 
\Im {\cal G}_0(\omega)
\end{eqnarray} 
where $f(\omega)$ is again the Fermi function. From 
\eqref{C10}  and \eqref{G65},\eqref{G66}  
the $f$-electron Green functions 
of the original Falicov-Kimball model can be found.  
\end{appendix}


\begin{thebibliography}{100}
\bibitem{falicov_kimball}  L.~M.~Falicov and J.~C.~Kimball, Phys. Rev. Lett.
{\bf 22}, 997 (1969)
\bibitem{freericks.98} 
J.~K.~Freericks and V.~Zlati\'c, Phys. Rev. B {\bf 58}, 322 (1998).
\bibitem{freericks.03} J.~K.~Freericks and V.~Zlatic, Rev. Mod. Phys. \textbf{75},
  1333 (2003).
\bibitem{brandt.89} U.~Brandt and C.~Mielsch, Z.~Phys.~B: Condens.~Matter 
\textbf{75}, 365 (1989).
\bibitem{dongen.90} P.~G.~J.~van Dongen and D.~Vollhardt,
  Phys.~Rev.~Lett. \textbf{65},
1663 (1990).
\bibitem{kennedy.86} T.~Kennedy and E.~H.~Lieb, Physica A \textbf{138}, 
320 (1986).
\bibitem{lieb.86} E.~H.~Lieb, Physica A \textbf{140}, 240 (1986).
\bibitem{sykora.05} S.~Sykora, A.~H\"ubsch, and K.~W.~Becker,
  Eur. Phys. J. B \textbf{51}, 181 (2006).
\bibitem{becker} K.~W.~Becker, A.~H\"ubsch, and T.~Sommer, Phys. Rev. B 
  {\bf 66}, 235115 (2002).
\bibitem{kohn.64} W.~Kohn, Phys. Rev. A \textbf{133}, 171 (1964).
\bibitem{schrieffer-wolff} J.~R.~Schrieffer, P.~A.~Wolff,
  Phys. Rev. \textbf{149}, 491 (1966).
\bibitem{anderson} P.~W.~Anderson, J. Phys. C: {\it Solid State Phys.}
  \textbf{3}, 2436 (1970). 
\bibitem{wegner} F.~Wegner, Ann. Phys. (Leipzig) {\bf 3}, 77 (1994).
\bibitem{wilson} S.~D.~G{\l}azek and K.G.~Wilson, Phys. Rev. D {\bf 48}, 5863
  (1993); S.~D.~G{\l}azek and K.G.~Wilson, Phys. Rev. D {\bf 49}, 4214 (1994).
\bibitem{Ott} H.~R.~Ott and Z.~Fisk, in \textit{Handbook on the Physics and
Chemistry of the Actinides}, edited by A: J.~Freeman and G.~H.~Lander
(Elsevier, Amsterdam, 1987).
\bibitem{QMC} P.~Farkasovsky, Phys. Rev. B \textbf{54}, 7865 (1996).
\bibitem{farkasovsky.03} P.~Farkasovsky, Int. J. Mod. Phys. B \textbf{17},
  4897 (2003).
\bibitem{nagaosa.98} in N.~Nagaosa, {\it Quantum Field Theory in Strongly
Correlated Electronic Systems} Springer-Verlag Berlin  
Heidelberg New York, 1998.   
\bibitem{maska.05} M.~M.~Maska and K.~Czajka, Phys.~Rev.~B \textbf{74}, 035109 (2006).
\bibitem{fick.90} in {\it The Quantum Statistics of Dynamic Processes},
E.~Fick and G.~Sauermann, Springer series in Solid-State Sciences 86,
1990.
\bibitem{Lyzwa.93} R.~Lyzwa, Physica A \textbf{192}, 
231 (1993).

\end{thebibliography}
\end{document}